\documentclass[a4paper,fleqn,usenatbib]{mnras}
\usepackage{graphicx}	
\usepackage{amsmath}	
\usepackage{amssymb}	
\usepackage{comment}

\def\lsun{{\rm L_{\odot}}}
\def\msun{{\rm M_{\odot}}}

\def\be{\begin{equation}}
\def\ee{\end{equation}}

\def\del#1{{}}
\newcommand\mearth{{\,{\rm M}_{\oplus}}}
\newcommand\mj{{\,{\rm M}_{\rm J}}}
\newcommand\rj{{\,{\rm R}_{\rm J}}}
\newcommand\MSunPerYear{~${\rm M_{\odot}}$~yr$^{-1}$\,}

\newcommand{\bref}[1]{{ #1}}

\title[Nayakshin and Elbakyan]{On the origin of accretion bursts in FUORs}

\author[Nayakshin \& Elbakyan]{Sergei Nayakshin$^1$\thanks{sn85@le.ac.uk} and 
Vardan Elbakyan$^1$ \\
$^{1}$School of Physics and Astronomy, University of
  Leicester, Leicester LE1 7RH, UK. 
}

\date{Accepted XXX. Received YYY; in original form ZZZ}

\pubyear{2022}

\begin{document}
\label{firstpage}
\pagerange{\pageref{firstpage}--\pageref{lastpage}}
\maketitle

\begin{abstract}
Accretion luminosity of young star FU Ori increased from undetectable levels to hundreds of $\lsun$ in 1937 and remains nearly as high at the present time. In a recent paper we showed how Extreme Evaporation (EE) of a young gas giant planet that migrated to a $\sim 10$ day orbit around the star may power FU Ori. However, our model assumed a power-law mass-radius relation for the evaporating planet. Here we employ a stellar evolution code to model mass losing planets. We find that adiabatic planets expand rapidly, which results in runaway FUOR bursts. Super-adiabatic planets contract while losing mass; their outbursts are dimming with time. Long steadily declining bursts such as FU Ori require  relatively fine tuned internal planetary structure, which may be rare. More commonly we find that super-adiabatic planets contract too rapidly and their EE falters, leading to FUOR burst stutter. This stutter allows a single planet to produce many short repeating bursts, which may be relevant to bursts observed in V346 Nor, V899, V1647. We compute broad band spectra of our best fitting scenario for FU Ori. Since the outburst is triggered behind the planet location, the mid-IR emission rises many months before the optical, similar to bursts in {\em Gaia}-17bpi and {\em Gaia}-18dvy. We show that in outbursts powered by the classic thermal instability, mid-IR lags the optical, whereas the dead zone activation models predict mid-IR light precede the optical burst by many years to decades. We comment on the stellar flyby scenario for FU Ori.
\end{abstract}


\section{Introduction}

During FU Ori outbursts \citep[FUOR hereafter;]{HK96,Fischer-PPVII}, protostars  accrete gas at astonishingly high rates of $\dot M \sim (10^{-6}-10^{-4})\, \msun$~year$^{-1}$ \citep{HartmannK85-FUORs,Zhu09-FUOri-obs}. There is a number of models for triggering these events \citep{AudardEtal14}. In the well known Hydrogen ionisation instability scenario \citep{Bell94}, also known as Thermal Instability (TI), the inner $\sim 0.1$~AU of the disc keeps switching between the stable low $\dot M$ and the high $\dot M$ solution branches. On the former, the disc midplane temperature is $T_{\rm d} \lesssim 3,000$~K and Hydrogen is neutral, whereas on the high branch $T_{\rm d} \gtrsim 30,000$~K, and Hydrogen is ionised. Disc viscosity is low on the former and high on the latter branches, leading to \bref{quiescent} $\dot M_{\rm low} \lesssim (10^{-8}-10^{-7})$ \MSunPerYear and \bref{outburst} $\dot M_{\rm high}\gtrsim 10^{-5}$ \MSunPerYear. However, to match outburst duration and accreted mass budget, \cite{Bell94} required disc viscosity to be two orders of magnitude lower than generally accepted based on both observations \citep[e.g.,][]{Lasota01-Review,Hameury-20-review} and numerical simulations \citep[e.g.,][]{Hirose15,Scepi-18-TI-alpha}.



\cite{LodatoClarke04} presented a TI-planet scenario for FUORs, in which a massive planet embedded in the disc \bref{opens a deep gap in the disc and leads to banking up of significant excess material behind its orbit. In this case the outbursts may start in this excess material rather than at $\sim 2-3$ stellar radii \citep[as happens in the absence of a planet;][]{Bell94}}. However, there is  a serious planet budget problem with this scenario. FUOR events are believed to occur roughly $\sim 10$ times per star \citep{HK96}, whereas only $\sim 1$\% of FGK stars host hot jupiters \citep[cf. Fig. 9 in][]{SanterneEtal15}.

\cite{NayakshinLodato12} argued that massive young planets can be fluffy enough to be tidally disrupted at separations $\sim 0.1$~AU. The matterial released by the planet is deposited in the protoplanetary disc and \bref{accretes onto the star quickly,} powering an accretion burst. \bref{In this picture most of very young hot jupiters are destroyed in FUOR bursts, so very few of them survive to be observed around mature stars.} This model is \bref{very closely related} to the influential scenario for episodic accretion proposed by \cite{VB06,VB10,VB15}. \bref{Large scale 2D simulations by these authors show} that planets born by gravitational instability in massive young discs at $\sim 100$~AU migrate into the inner $\sim 10$ AU of the disc very rapidly. If accreted by the star these planets account for both the frequency and the mass budget of FUORs qualitatively well. The \cite{NayakshinLodato12} calculations describe how this planet accretion process \bref{works if the planets are tidally disrupted} on scales unresolved in the large scale 2D simulations. Unfortunately, this scenario fails decisively in terms of outburst light curves. Planet tidal disruptions result in outbursts that are too short and too bright by $\sim$ two orders of magnitude. \cite{Armitage15-review}, his \S 6, also notes that it is hard to see how gas giant planets can be as extended as $\sim 40 \rj$ to be tidally disrupted at $\sim 0.1$~AU.

\cite{Nayakshin-23-FUOR} (hereafter paper I) \bref{pointed out a number of reasons why a massive planet may be key to understanding FU Ori:} (i) Absorption lines profiles and quasi-periodic photometric variability of FU Ori indicate the presence of a hot spot in the disc at the distance of $\sim 0.08$~AU from the star \citep[e.g.,][]{PowellEtal12,Siwak21-FUOri-QPOs}, with the period steady for two decades now \citep[see also][]{Herbig03-FUOR}. (ii) Interferrometric observations of FU Ori by \cite{2022Lykou} showed that the active disc feeding the star for nearly a century at the rate well above $10^{-5}$\MSunPerYear extends to the radius of only $\sim 0.3$~AU\footnote{In retrospect, this value is probably very much consistent with earlier findings from SED disc modelling by \cite{ZhuEtal07,Zhu08-FUOri}. At the time their modelling was done, FU Ori distance, mass, and disc inclinations were all assumed to be significantly different from the better constrained modern values given in \cite{2022Lykou}. Using the latter, and requiring the active disc to stop at exactly the same effective temperature as in \cite{Zhu08-FUOri} we rescale their value of $R_{\rm act} = 0.58$~AU to 0.296 AU.}. This is paradoxical unless there is a source of mass hidden inside of this tiny region. The outward viscous spreading of material from $a=0.08$ naturally results in an outer bright disc edge at $\sim 0.3$~AU. (iii) \bref{The slow monotonic decline of FU Ori brightness from 1937 till now is paradoxical since the disc is expected to exhibit TI. However, TI was found to disappear if the source of  matter feeding FU Ori is not the outer disc but a planet at 0.08 AU.}

We then proceeded to describe and characterise a previously missed planet-disc interaction process, named Extreme Evaporation (EE). It was shown that dusty GI planets have radii as large as $R_{\rm p} \sim (10-20)R_{\rm J}$ at the age of class I YSOs. When such a planet migrates into the inner $ 0.1$~AU of the disc, it is exposed to midplane disc temperatures $T_{\rm d} \gtrsim 30,000$~K during TI outburst episodes. As the planet outer layers heat up to these extreme temperatures, they become unbound from the planet. As in \cite{NayakshinLodato12} model,  mass lost by the planet is  deposited in the protoplanetary disc; the difference is in how the planet mass is lost. While tidal disruption of a gas giant planet is usually a runaway (almost annihilation-like) event, EE is a self-limiting quasi-steady process. Using a time-dependent disc with an embedded planet model we have shown in paper I that this TI-EE version of the model accounts for a number of observed features of FU Ori and interferrometry constraints.

In this paper we address two important outstanding questions. First, planet radius evolution during mass loss via EE was found (see fig. 15 in paper I) to be crucial in governing outburst properties, yet a simple power-law mass-radius ($M-R$) relation was assumed. Second, in paper I we focused on the time evolution of the accretion rate onto the star as a proxy for disc brightness; however this leaves behind the more observationally pertinent question of how the spectrum of the system evolves in various filters.

The paper is structured as following. A brief account of computational methods is given in \S 2. In \S \ref{sec:toy_expectations} we explain in simple terms why the internal structure of the planet is key to the planet-sourced TI-EE scenario of FUORs. In \S \ref{sec:M_R_linS} stellar evolution code MESA is used to explore $M-R$ relation for a standard convective versus linearly super-adiabatic planet models; the differences in resulting accretion bursts is presented in \S \ref{sec:Bursts_linS}. In \S \ref{sec:M_R_expS} we experiment with exponential super-adiabatic planets which have inert solid cores. In \S \ref{sec:SED} we compute the Spectral Energy Distribution (SED) of time-dependent discs in model outbursts for three competing models: TI (classical planet-free thermal instability), TI-EE, and the MRI activation in the dead zone \citep[DZ-MRI hereafter; ][]{ArmitageEtal01}.  \bref{We also discuss the stellar flyby scenario for FU Ori in \S \ref{sec:external_perturbers}.} An appendix investigates the claim made in paper I that TI cannot be naturally suppressed in the classical DZ-MRI model of FU Ori.

\section{Methods}

\subsection{Disc-planet evolutionary code DEO}\label{sec:DEO}

DEO (Disc with an Embedded Object) is a time-dependent 1D code for evolving a \cite{Shakura73} vertically averaged disc model with an (optionally) embedded planet in it \citep{NayakshinLodato12,Nayakshin22-ALMA-CA}. We only give a brief account of it here (cf. paper I for more detail). The disc surface density $\Sigma$ is evolved according to
\begin{equation}
\begin{split}
    \frac{\partial\Sigma}{\partial t} = \frac{3}{R} \frac{\partial}{\partial R} \left[ R^{1/2} \frac{\partial}{\partial R} \left(R^{1/2}\nu \Sigma\right) \right] - \\
    - \frac{1}{R} \frac{\partial}{\partial R} \left(2\Omega^{-1} \lambda \Sigma\right) + 
    \frac{\dot{M}_{\rm p}}{2\pi R} D(R - a)\;.
\end{split}
\label{dSigma_dt0}
\end{equation}
In this equation $\Omega$ is the Keplerian angular velocity, $\nu=\alpha c_{\rm{s}} H$ is the kinematic viscosity, $H$ is the disc vertical scale height, and $c_{\rm{s}}$ is the disc sound speed. The exchange of angular momentum between the planet and the disc is described by the second term on the right hand side of the equation \citep[expressions for $\lambda$ are given in][]{Nayakshin22-ALMA-CA}, whereas the last term describes the injection of mass into the disc if/when the planet loses mass at the rate $\dot M_{\rm p}$. The function $D(R-a)$ is a narrow Gaussian normalised to yield $\int_0^\infty 2\pi R D(R) dR = 1$. The planet-star separation, $a$, evolves in a way explicitly conserving angular momentum of the disc-planet system. Mass is deposited into our disc at the feeding rate, $\dot M_{\rm feed}$, close to the outer disc boundary.

Planet evaporation is driven by exchange of heat between the disc and the planet. The disc midplane temperature, $T_{\rm d}$, and the planet radius, $R_{\rm p}$, are the primary parameters determining the rate of planet mass loss, $\dot M_{\rm p}$. When $R_{\rm p} $ is less than the Bondi radius,
\begin{equation}
       R_{\rm B} = \frac{G M_{\rm p}}{2 c_s^2} \approx 12\, R_{\rm J} \, \frac{M_{\rm p}}{5\mj}\, \left(\frac{T_d}{3\times 10^4}\right)^{-1}\;,
    \label{Rbondi}
\end{equation}
the outer layers of the planet remain marginally bound to it, and so mass loss proceeds via the relatively weak Bondi wind \citep[``boil off"; ][]{OwenWu-16-boil-off} solution. However, when $R_{\rm p} > R_{\rm B}$, the outer heated layers of the planet are unbound and are removed rapidly at the EE rate that is well approximated by
\begin{equation}
    \dot M_{\rm EE} = 1.6\times 10^{-5} 
    \frac{\msun}{\hbox{year}} \left(\frac{T_{\rm d}}{3 \times 10^4 \hbox{K}} \right)^{2.2}\, \left(\frac{R_{\rm p}}{10 R_{\rm J}}\right)^{3/2}\;.
    \label{Mdot_fit}
\end{equation}
Note that disc TI is needed to trigger EE in this picture, and only if $\dot M_{\rm EE}$ is larger than $\dot M$ that the disc would have in the  absence of the planet does the planet-disc system enter the self-sustained ``planet-sourced" mode. In this case the  planet plays the role of a mass losing secondary star in a binary system; this can last for tens to hundreds of years.


\subsection{Planet evolution and mass-radius relation}\label{sec:MESA_models}

Here we use MESA stellar evolution code \citep{PaxtonEtal11,Paxton13-Mesa-2} to model planet evolution. All of our calculations are performed for uniform composition planets with metallicity $Z = 0.02$. We use the standard MESA opacity for brown dwarfs and planets \citep{Freedman08-opacity} supplemented by silicates dust opacity. The dust opacity is given by
\begin{equation}
    \kappa_{\rm d} = 3 \;\text{cm}^2\; \text{g}^{-1}\; \left(\frac{T}{10^3 \text{ K}} \right) \; f_{\rm melt}(T, \rho)
    \;,
    \label{kappa_d}
\end{equation}
where the function $f_{\rm melt}$ approximates the effect of grain melting \citep[following][]{Kuiper10-RHD} via
\begin{equation}
    f_{\rm melt}(T, \rho) = \frac{1}{2} - \frac{1}{\pi} \arctan\left(\frac{T-T_{\rm melt}}{100 \text{ K}}\right)
\end{equation}
with melting temperature depending on gas \bref{density} as
\begin{equation}
    T_{\rm melt} = 2000\; \text{K } \rho^{0.0195}\;.
\end{equation}
Note that this assumes Solar metallicity of grains in the atmosphere of the planet, and that $f_{\rm melt}= 1$ for $T \ll T_{\rm melt}$.

In paper I (\S 7) we constructed models in which a planet started far from the inner disc region with a large initial radius, $R_{\rm p}$. We then computed $R_{\rm p}$ evolution together with planet migration in the protoplanetary disc. We neglected planet mass loss in the Bondi wind regime, and turned on EE (eq. \ref{Mdot_fit}) once $R_{\rm B}$ fell below $R_{\rm p}$. At that point our planet evolution simply assumed that $R_{\rm p}$ remained constant or dependent on the changing $M_{\rm p}$ as a power-law. Simulation labelled M1 in Paper I reproduced many traits of the observed FU Ori outburst. In this simulation, the planet had initial mass $M_{\rm p0}=6\mj$ and radius $\approx 14 R_J$ at the commencement of EE. The resulting mass loss rate varied during the burst from the peak of $4\times 10^{-5}\msun$~year$^{-1}$ to $\sim 3\times 10^{-5}\msun$~year$^{-1}$, {\em under the assumption } $R_{\rm p}=$~const. 

Here we relax this assumption. To limit the parameter space that we need to investigate, we continue with the planet of $M_{\rm p0}=6\mj$ and radius $\approx 14 R_J$ at the commencement of EE, but here we use MESA to study planet radius evolution when the planet loses mass vigorously. We do not model the outflow region directly as this requires explicit hydrodynamics with very short time steps. Instead we prescribe a mass loss rate $\dot M_{\rm p}$ and follow the standard approach to mass loss in MESA (mass is removed from the outer regions of the planet as described in the MESA instrument papers). 

\begin{figure}
\includegraphics[width=0.99\columnwidth]{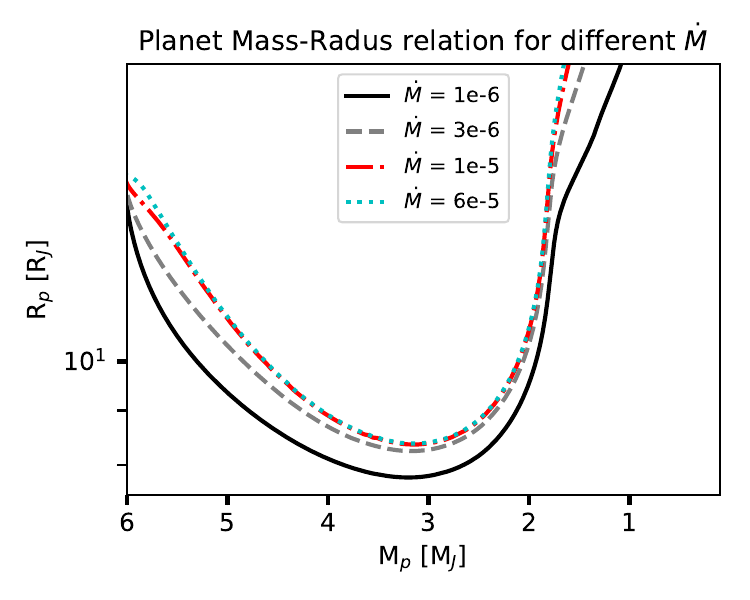}
\caption{Planet radii versus planet mass for the same initial planet structure but different planet mass loss rates,  $\dot M$. This shows that $M-R$ relation is nearly independent of $\dot M_{\rm p}$ for mass loss rates exceeding a few $10^{-6}\msun$~year$^{-1}$. } 
\label{fig:No_Mdot_dependency}
\end{figure}

One fortunate result simplifies the otherwise non trivial (and not yet achieved) on-the-fly coupling of our disc code DEO with MESA. Fig. \ref{fig:No_Mdot_dependency} shows the mass-radius relation for several MESA runs that start with an identical planet internal structure (model $S_{\rm min}=12$ described in \S \ref{sec:M_R_linS}), but different $\dot M_{\rm p}$. We see that the mass-radius relation $R(M)$ is independent of the exact value of $\dot M_{\rm p}$ as long as it exceeds $\sim 3\times 10^{-6}\msun$~year$^{-1}$. This is to be expected\footnote{At very high $\dot M_{\rm p}$ the planet has no time to shuffle energy between different layers by either convection or radiation. Qualitatively, radiative losses are not important during mass loss if mass loss time scale $M_{\rm p}/\dot M_{\rm p} \ll t_{\rm KH}$, the Kelvin-Helmholz timescale. For the problem at hand, $t_{\rm kh}\sim O(10^4)$ years as our planets are contracting as they are migrating, and their migration time scale is of this order. Therefore, by order of magnitude, for mass loss rates $\dot M_{\rm p} \gg M_{\rm p}/t_{\rm kh}\sim 10^{-6} $\MSunPerYear planet $M-R$ relation should be independent of $\dot M_{\rm p}$.}.  Only for $\dot M_{\rm p} = 10^{-6}\msun$~year$^{-1}$ there is a $\sim 10$\% difference in $R_{\rm p}$ at a given $M_{\rm p}$. Such low $\dot M_{\rm p}$ are of limited interest for us here, so we fix $\dot M_{\rm p} = 3\times 10^{-5}\msun$~year$^{-1}$ for all our MESA planet mass loss calculations below. \bref{We emphasise that this approach is accurate within  about a few \% as per Fig. \ref{fig:No_Mdot_dependency}.}

\section{Coreless planets with a linear entropy function}\label{sec:M_R_linear_entropy}

\subsection{Why planet internal structure matters, and why its uncertain}\label{sec:toy_expectations}

In this paper we assume that FU Ori accretion rate varied with time approximately as
\begin{equation}
    \dot M \approx 5\times 10^{-5} \; \frac{\msun}{ \hbox{yr}} \; e^{-t/t_{\rm e}}\;,
    \label{dotM_obs}
\end{equation}
where $t_{\rm e} = 73$ years and $t=0$ corresponds to the beginning of the burst in year 1937 \citep{Herbig66-FUORi}. This equation is consistent with the current $\dot M$ in FU Ori  \citep[the ``ALMA" model in table 1 of][]{2022Lykou} and the dimming rate of 0.015 magnitude per year \citep{KenyonK-2000-FUOR}. If $\dot M$ in FU Ori is approximately equal to $\dot M_{\rm EE}$, then eq. \ref{Mdot_fit} suggests that planet radius decreased with time approximately exponentially with e-folding time of $\sim 120$ years. 

We start with qualitative ideas. Hydrogen in very young post collapse planets is nearly completely ionised, so we may hope that a toy ``ideal polytrope" planet with a constant mean molecular weight, $\mu$, a constant specific entropy, $S$, and an adiabatic equation of state (EOS), pressure $P= K\rho^\gamma$, with $\gamma = 5/3$, is appropriate.  For such a planet \citep[\S 19 in][]{Kippenhahn13-book}
\begin{equation}
    R_{\rm p} \propto K M_{\rm p}^{-\xi_{\rm p}} \;,
    \label{M-R-ideal}
\end{equation}
where $\xi_{\rm p} = 1/3$, so $R_{\rm p}$ increases as $M_{\rm p}$ drops. Such a planet expands as it loses mass, so its FU Ori outburst  would brighten with time in contrast to eq. \ref{dotM_obs}.

\begin{figure}
\includegraphics[width=0.99\columnwidth]{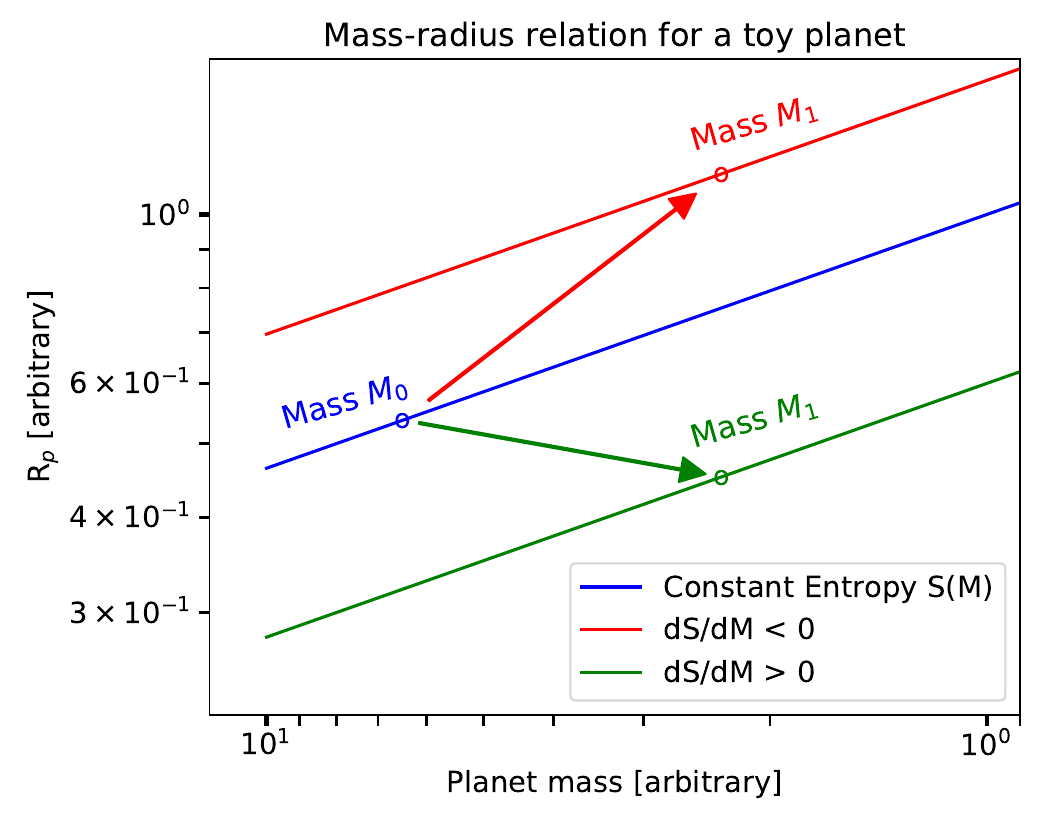}
\caption{Planet radius versus mass for a toy planet in which entropy varies with enclosed mass. Super-adiabatic planets (positive $dS/dM$) can contract while losing mass. See \S \ref{sec:toy_expectations} for detail.}
\label{fig:Toy_model_planet}
\end{figure}

In a more general case entropy $S$ varies as a function of enclosed mass $M$ within the planet. Fig. \ref{fig:Toy_model_planet} is a sketch of the mass-radius relation for such planets (note that $M_{\rm p}$ decreases along the horizontal axis). Consider three planets with the same initial mass, $M_0$, and radius, $R_0$, but with different internal entropy profiles. Let the planets lose mass so rapidly that neither radiation nor convection are able to transfer energy between different layers in the planet. The planets in Fig. \ref{fig:Toy_model_planet} start at the the blue dot but evolve differently:

\begin{enumerate}
    \item In the ``blue" planet the entropy is constant with $M$. As it loses mass it moves along the blue line towards the top right corner of the figure, following eq. \ref{M-R-ideal} with $\xi_{\rm p} = 1/3$. 
    \item In the ``red" planet the entropy is higher in the centre. For the sake of simplicity let it be constant within enclosed mass $M_1$. When the planet mass becomes equal to $M_1$ its radius then is equal to that at the red point on the red curve. This planet mass-radius relation is  steeper, $\xi_{\rm p} > 1/3$.  
    \item Finally, in the ``green" planet the entropy is lower in the centre. When this planet mass decreases to $M_1$ it arrives at the green point. This planet mass-radius relation is shallower, so $\xi_{\rm p} < 1/3$, and may be negative. 
\end{enumerate}

The internal structure of GI planets with age $\sim O(10^4)$ years is currently uncertain. The planets are born at separations of $\sim O(100)$~AU as pre-collapse molecular Hydrogen dominated clumps. They contract until central temperature $T_{\rm c}\sim 2000$~K is reached, at which point H$_2$ dissociates and the planet collapses into the post-collapse configuration \citep{Bodenheimer74}. This is similar to the collapse of first cores into the second cores in star formation \citep{Larson69,Masunaga00} but occurs in a disc. No detailed calculation of such a collapse has been performed to date, to the best of our knowledge. However, \cite{Bate-14-2nd-core} perform 3D radiation transfer MHD simulations of a rotating $1\msun$ cloud collapse down to stellar core densities. While not performed for protoplanetary collapse, their simulations address the properties of the second cores formed early on in the collapse. They find that the sizes of cores with masses up to $20 \mj$ are $\sim 30 \rj$ and that they have entropy profiles increasing outward, with the minimum entropy of $S\approx 14 k_b/m_p$. Similar entropy values and even larger core sizes are found in 2D radiative hydrodynamics calculations of \cite{2Bhandare-20-2nd-cores}, e.g., see their Figs. 3 and 8. 

This large initial entropy of a protoplanet will be reduced by radiative cooling, however entropy of the outer planetary layers can also be increased by energy deposition due to tides, irradiation, or other effects \citep[e.g.,][]{Bodenheimer-03-Tidal-Heating,Jackson08-Tidal-Heaing,GinzburgSari15-inflated-hotJ} when the planet migrates into the innermost disc. Additionally, gas accreted by the planet would initially inherit the disc gas entropy, which is significantly higher than that of post-collapse planets. Simulations show that GI planets can be born in metal-enriched regions of the disc \citep[e.g.,][]{BoleyDurisen10} and they accrete pebbles rapidly \citep{HN18,Vorobyov-Elbakyan-19,Baehr19-pebble-accretion}. This deposition of heavy elements into the planet is certain to create non-uniform and non-adiabatic planets \citep[e.g.,][]{VallettaHelled-20-high-Z,Ormel-21_pebbles_interior}. This motivates us to explore different assumptions about planet internal structure in \S \ref{sec:M_R_linS} and \S \ref{sec:M_R_expS}.

In the interest of connecting to previous literature we point out that the uncertainties in the initial structure of GI planets are probably less important after the disc is dissipated (after $\sim 3$ Myr). Once planet ``harassment" via disc interactions stops the planet will cool and should become nearly  fully convective and therefore very close to being adiabatic \citep{GraboskeEtal75,BodenheimerEtal80,WuchterlEtal00,HelledEtal08}. This is why in stellar evolution codes the planets are usually initialised as constant entropy spheres \citep[e.g.,][]{BurrowsEtal97,SpiegelBurrows12,Paxton13-Mesa-2}. The justification for this approach is that the initial contraction is ``rapid"(e.g., cf. fig. 16 in paper I) and in $\sim 10^6$ years the exact initial conditions are forgotten.




\subsection{Planet response to mass loss}\label{sec:M_R_linS}

Here we continue to focus on planets with initial mass and radius $M_{\rm p0}=6\mj$ and $R_{\rm p0} = 14 R_{\rm J}$, respectively, as per model M1 from paper I. We want to know how such planets respond to vigorous mass loss  for different internal structure. Fig. \ref{fig:Radius_vs_t0} shows the results of our MESA calculations with solid curves. We explain these curves below. The bottom and top horisontal axises show time, counted from commencement of mass loss, and planet mass, respectively.

The ideal polytrope $M-R$ relation (eq. \ref{M-R-ideal}) is depicted in Fig. \ref{fig:Radius_vs_t0} with the  violet dash-dotted curve. The shaded region between the dashed cyan and the dotted green curves is important for the interpretation of the resulting FU Ori model outbursts in \S \ref{sec:Bursts_linS}. The upper curve is the planet Hill radius at separation of 0.08 AU. If the planet swells up and crosses the dotted green curve, it is destroyed tidally \citep[as in][]{NayakshinLodato12} very rapidly, resulting in a short and very powerful burst very unlike FU Ori. The cyan dashed curve in Fig. \ref{fig:Radius_vs_t0} is the Bondi radius (eq. \ref{Rbondi} for fixed $T_d = 3.75\times 10^4$ K), which drops together with $M_{\rm p}$. If $R_{\rm p}$ drops below $R_{\rm B}$ then EE process stops. To remain in the EE regime the planet must be in the shaded area.


\subsubsection{Standard planet}\label{sec:standard_planet}

The black curve is the $M-R$ relation of the actual MESA model we used for simulation M1 from paper I. We label this model ``standard" since except for addition of dust opacity in the planet envelope it is the one obtained with the default initial conditions for young post-collapse planets in MESA \citep[procedure {\it create\_initial\_model } in][]{Paxton13-Mesa-2}. The planet was initialised as a constant entropy sphere with radius $\approx 30 \rj$. It was then cooling and contracting for about 20 thousand years while migrating from the outer disc to $R\approx 0.08$~AU where it entered the EE regime (cf. figs. 16, 22 and 23 from paper I). 

Fig. \ref{fig:Radius_vs_t0} shows that the standard planet expands very rapidly, roughly following the $\xi_{\rm p} \sim 2$ track, much steeper than the ideal polytrope (violet dash-dot curve). This may appear surprising since the entropy profile $S(M)$ within the standard planet is very close to a constant, linearly decreasing with $M$ from the maximum of $S(0)\approx 14.25$ in units of $k_b/m_p$ in the centre to the minimum of $S(M=M_{\rm p}) \sim 13.9$ (there is an upturn in $S$ close to the planet surface but this involves a very small fraction of mass and cannot affect the evolution of $R_{\rm p}$ strongly). The entropy profile decreasing with $M$ is a natural outcome of radiative cooling for a planet that is initialised adiabatic. The planet cools and contracts by losing energy from its surface. The outer layers thus lose entropy first, however convection maintains a {\em nearly} constant $S(M)$.  

The main reason for the rapid expansion of the standard planet is recombination of Hydrogen which releases a significant amount of energy not taken into account in the toy model planet in eq. \ref{M-R-ideal}. At $t=0$, the mean H ionisation fraction in the ``standard" planet is about 50 \%, but this falls very rapidly as the planet expands and its mean temperature drops. The line $R_{\rm p} = R_{\rm H}$ is crossed  due to this rapid expansion at $t=40$ years, when the planet mass is $\sim 4.8\mj$.

\subsubsection{Super adiabatic planets}\label{sec:super_adiabatic_lin}

\begin{figure}
\includegraphics[width=0.99\columnwidth]{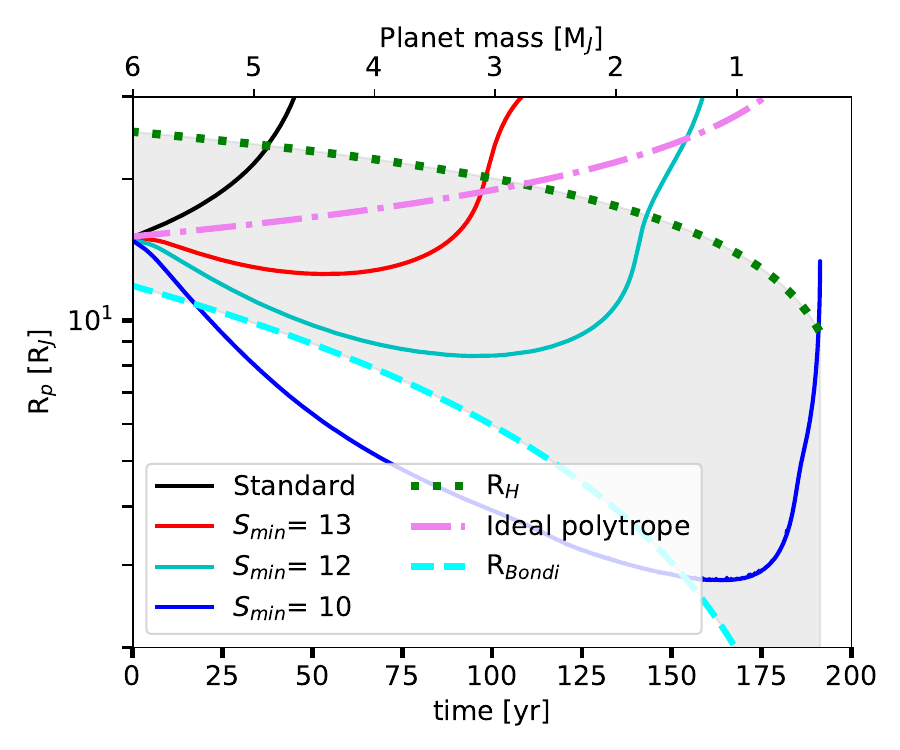}
\caption{Planet radii versus time for a selection of MESA model planets losing mass at rate  $\dot M_{\rm p} = 3\times 10^{-5}\msun$~year$^{-1}$ (solid curves). Depending on the internal structure, the planets either expand or contract. To be in the EE regime continuously, $R_{\rm p}$ must be in the shaded area. Planets leaving the area through the top are tidally destroyed; planets contracting too rapidly and falling below the cyan dashed curve stop losing mass and so FU Ori outbursts may switch off. See text in \S \ref{sec:M_R_linS} for detail.}
\label{fig:Radius_vs_t0}
\end{figure}

The rest of the solid curves in Fig. \ref{fig:Radius_vs_t0} are for {\em super-adiabatic} planets, that is, those with entropy $S(M)$ increasing outwards in an arbitrary postulated linear fashion:
\begin{equation}
    S(M) = S_{\rm min} + (S_{\rm max}-S_{\rm min}) \frac{M}{M_{\rm p0}}\;,
    \label{S_heat_linear}
\end{equation}
where $S_{\rm min}$ is the entropy in the planet centre, whereas $S_{\rm max}$ is entropy at its outer edge. To initialise such calculations with MESA we take the initial state of the standard planet and cool its centre while heating its outer layers until the entropy profile given by eq. \ref{S_heat_linear} is established. Since we aim for a planet initially having exactly the same $R_{\rm p}$, only one of the parameters $S_{\rm min}$ and $S_{\rm max}$ is independent; we chose to pick $S_{\rm min}$ and adjust $S_{\rm max}$ to yield $R_{\rm p} =14 R_{\rm J}$.

The legend in Fig. \ref{fig:Radius_vs_t0} lists $S_{\rm min}$ for the respective curves. We can see that super-adiabatic planets contract initially, as expected. However, at some point a minimum in $R_{\rm p}$ is reached, and contraction turns into expansion. The cooler the planet is in the centre (the smaller is $S_{\rm min}$), the later that minimum is reached. This eventual rapid expansion of the planet is once again strongly abetted by Hydrogen recombination.

\subsection{Resulting FUOR bursts}\label{sec:Bursts_linS}

Here we study how properties of EE bursts depend on the planet mass-radius relations we computed in \S \ref{sec:M_R_linS}. In this section we use the same disc model that we used for the model M1 in paper I and simply inject the planet into the disc on a circular orbit at $a= 0.08$~AU \bref{and hold its orbit fixed}. Our main focus here is how the disc-planet system behaves \bref{in response to}  how the planet radius evolves when $M_{\rm p}$ decreases as the EE burst proceeds. \bref{Planet migration will be self-consistently included in the calculations in \S \ref{sec:param_space}.}

\begin{figure*}
\includegraphics[width=0.99\textwidth]
{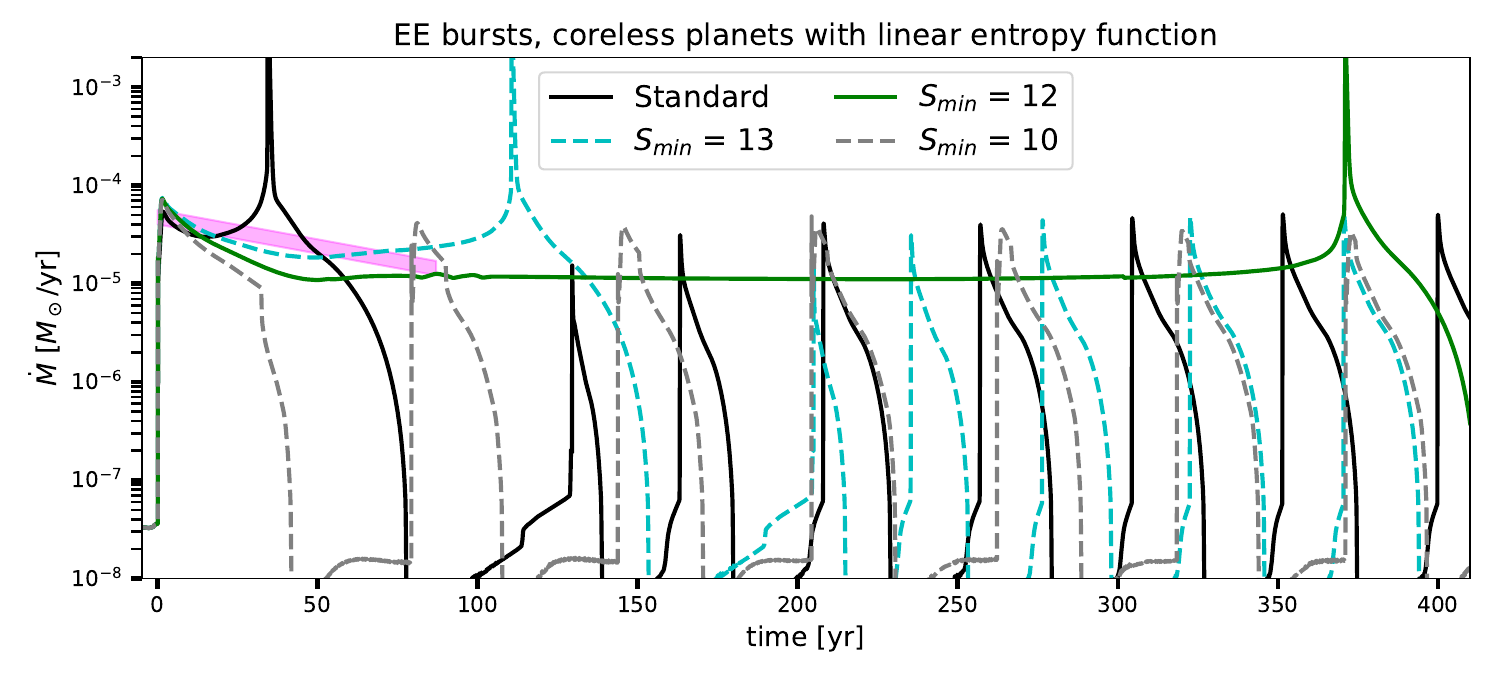}
\caption{Stellar mass accretion rates for EE bursts for planets with the same initial radius but different internal structure (see Fig. \ref{fig:Radius_vs_t0}). The standard MESA planet is nearly adiabatic and expands rapidly as the planet looses mass. $\dot M_{\rm EE}$ runs away and the planet is eventually tidally disrupted. Planets with smaller $S_{\rm min}$ are denser in the centre; they contract while losing mass and are more resilient to mass loss. The pink shaded area is the approximate observed FU Ori $\dot M$ evolution. See \S \ref{sec:Bursts_linS} for detail.}
\label{fig:FUOR_difS}
\end{figure*}

In Fig. \ref{fig:FUOR_difS} we show stellar accretion rate versus time after the planet starts to lose mass (the time axis is shifted to the beginning of the burst) for the four MESA model planets from Fig. \ref{fig:Radius_vs_t0}. We do not show planet mass loss rates for brevity and clarity of the figure. The green thick line is a power-law fit to $\dot M$ evolution of FU Ori given by eq. \ref{dotM_obs}.

The ``standard planet" model (solid curve) results in an accretion burst that  declines slightly during the first $\sim 15$ years but then runs away to an immense peak at $t\approx 36$ years. The initial decline is not due to planet radius evolution, rather this is simply a drop from the initial peak powered by both the planet and the disc TI burst.  Since the standard planet  expands with time,  $\dot M_{\rm EE}$ eventually increases, powering the corresponding increase in stellar $\dot M$. As foreseen in \S \ref{sec:M_R_linS}, this is a runaway process. Eventually the planet fills its Roche lobe ($R_{\rm p}$ exceeds $R_{\rm H}$) and the planet is disrupted tidally. 

As expected based on the arguments of \S \ref{sec:toy_expectations}, planets with $S(M)$ increasing outward produce accretion bursts with $\dot M$ decreasing more rapidly with time than for the standard planet. None of the three models however match FU Ori observations. For $S_{\rm min} = 13$, $\dot M$ decreases for about 45 years but then increases towards the eventual tidal disruption. $S_{\rm min} = 12$ model predicts a more rapid fall from the maximum brightness in FU Ori, and that the bursts would continue for almost 400 years in total when the planet is eventually disrupted. It is also somewhat inconsistent with the observations because FU Ori is continuing its declining trend currently \citep{2022Lykou}. 

Althrough this is not obvious from the figure, we note that the earlier the planet is disrupted tidally, the more powerful is the accretion outburst onto the star. This is because the planet mass tends to be larger.

The most centrally condensed planet, $S_{\rm min} = 10$ in Figs. \ref{fig:Radius_vs_t0} and \ref{fig:FUOR_difS}, contracts way too rapidly and is able to sustain the primary EE-dominated burst for only $\sim 30$ years. This is because the planet contracts and becomes smaller than the Bondi radius at that time, as expected based on the cyan dashed curve in Fig. \ref{fig:Radius_vs_t0}\footnote{Note that $R_{\rm B}$ in  Fig. \ref{fig:Radius_vs_t0} assumes a fixed disc temperature around the planet and sets the mean molecular weight to $0.63 m_{\rm p}$, whereas DEO calculations shown in Fig. \ref{fig:FUOR_difS} make no such approximations.}. 

Interestingly, multiple planet-powered bursts occur in the $S_{\rm min} = 10$ model. Thus, a planet that falls through the lower boundary of the shaded region in Fig. \ref{fig:Radius_vs_t0} is not at all lost to FUOR phenomenon but only the current burst.
When the next TI outburst occurs, the disc heats up again, and to a temperature higher than the one at the end of the primary EE burst. This means that $R_{\rm B}$ is smaller at the beginning of the next burst ($t \approx 80$~years) than it was at the end of the primary one. The $R_{\rm B} < R_{\rm p}$ criterion for EE process is satisfied again and hence a second episode of a sustained planet mass loss at an FU Ori-like rate results. The second episode is however shorter (only about 10 years long) than the previous one because the planet is now smaller. The third and so on episodes are shorter still, and so the planet may actually undergo very many smaller mass loss episodes. However, if planet migration was included in this calculation then such a planet could continue to migrate closer to the host star and could eventually be destroyed via either EE or TD, but this would typically occur $\sim O(10^3)$ years later.

\section{Planets with exponential entropy function and solid cores}\label{sec:M_R_expS}

\subsection{Mass radius relations}

As another class of planet internal structure models we consider the entropy function of the form
\begin{equation}
    S(M) = S_{\rm max} + (S_{\rm min}-S_{\rm max}) e^{-M/M_{\rm ent}}\;,
    \label{S_heat_exp}
\end{equation}
where $M_{\rm ent}$ is a ``low entropy core" of the planet, with $M_{\rm ent}$ a free parameter. By trial and error we found that there is a certain degeneracy of model results to the value of $S_{\rm min}$ and $M_{\rm ent}$, and so we only present here the results for $M_{\rm ent} =2 \mj$. 

We also add an inert solid core with mass $M_{\rm c}$ to the planet. A solid core in GI planets may form through grain growth and sedimentation \citep[e.g.,][]{McCreaWilliams65,Boss98,HelledEtal08,HS08,BoleyEtal10,Nayakshin10a,Nayakshin10b}, although such a core may be expected to be luminous \citep[e.g.,][]{HN19-planet-disruption} given its rather short growth time. We consider two contrasting values for  $M_{\rm c}$ here, one with a low mass core of $M_{\rm c} = 5\mearth$ and the other with a very massive one\footnote{Numerical simulations show that GI planets embedded in discs accrete pebbles rapidly and may therefore be metal rich compared to their host stars \citep{HN18,Vorobyov-Elbakyan-19,Baehr19-pebble-accretion}. They could therefore make very massive cores in principle.}, $M_{\rm c} = 65\mearth$. We assume that the core has a constant density of $\rho_{\rm core} = $~7.8 g/cm$^3$.

\bref{To refer to these exponential entropy profile models we shall use ``Exp-A-B" format where ``A" is the central entropy and "B" is the core mass.}

As in \S \ref{sec:M_R_linS}, before we begin our  MESA mass loss calculations, we adjust planet internal structure to the desired form (eq. \ref{S_heat_exp}).
We start with an initially adiabatic planet with a given central core mass and the value of $S_{\rm min}$ so that the planet is smaller than the desired radius $R_{\rm p}$. We then increase $S_{\rm max}$ sufficiently rapidly to preclude radiative cooling of the outer layers yet sufficiently slowly to allow the planet structure to adjust at sub-sonic speeds and follow eq. \ref{S_heat_exp} with an instantaneous value of $S_{\rm max}$. We keep increasing $S_{\rm max}$ until the planet expands to the desired value of $R_{\rm p} = 14\rj$.

\begin{figure*}
\includegraphics[width=0.99\textwidth]{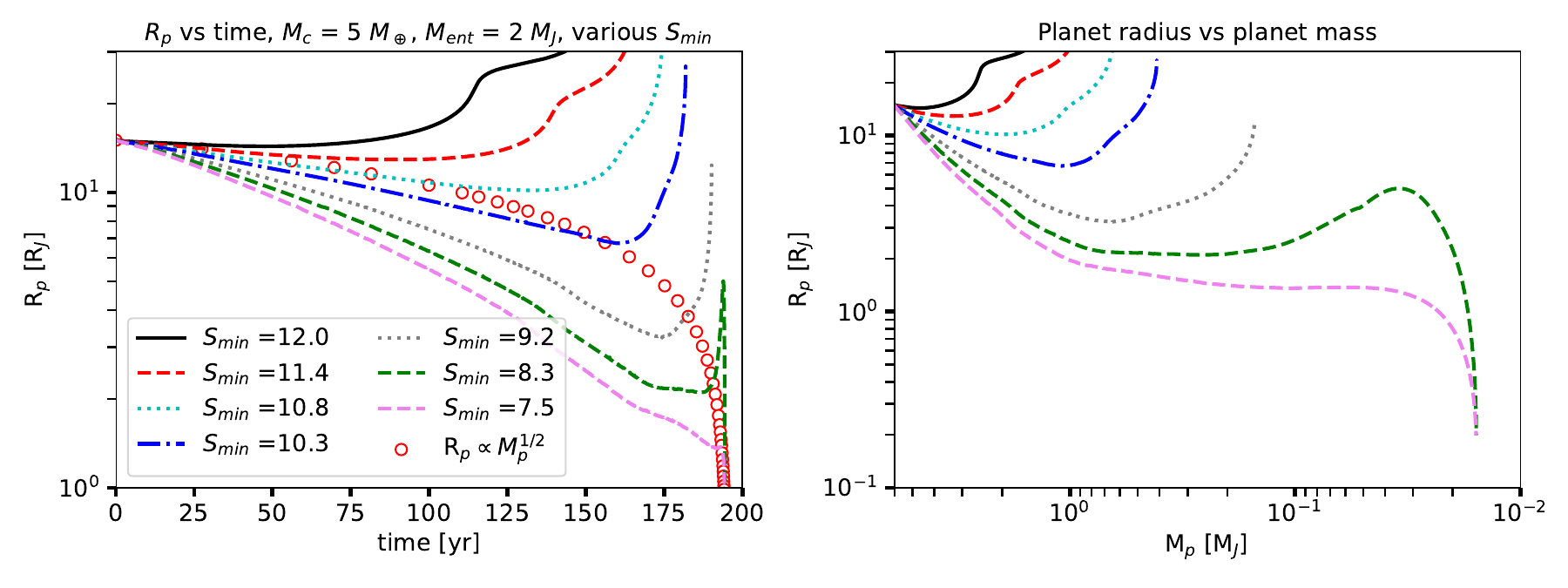}
\includegraphics[width=0.99\textwidth]{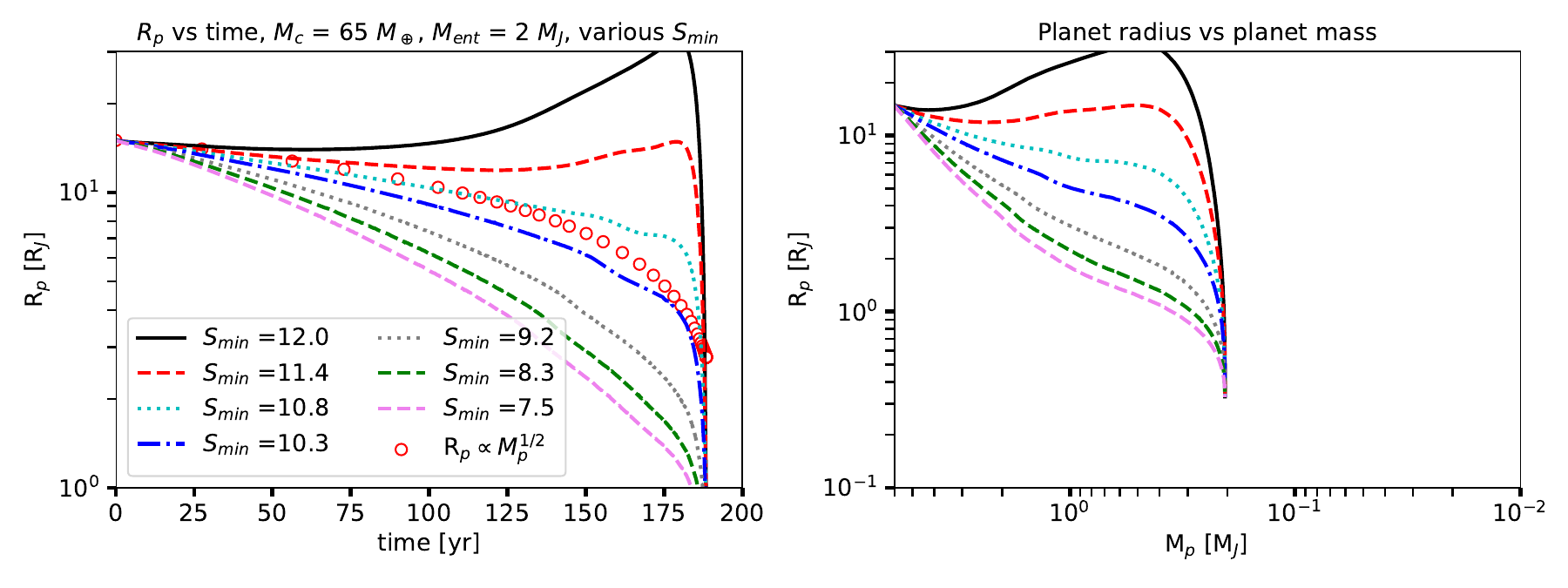}
\caption{{\em Left panels:} Planet radius vs time for planets with exponential entropy function (\S \ref{sec:M_R_expS}) losing mass at the rate $\dot M = 3\times 10^{-5}$\MSunPerYear. {\em Right panels:} Same results but plotted versus planet mass.} 
\label{fig:M_R_SA_with_core}
\end{figure*}

We computed mass-radius relation for mass-losing planets with $S_{\rm min}$ from $7.5$ to 12, and plot the results in Fig. \ref{fig:M_R_SA_with_core}. The figure also shows a simple power-law mass radius relation with $\xi_{\rm p} = 0.5$  with the open red circles for comparison.

Focusing first on the top panels, $M_{\rm c} = 5\mearth$, we note that the mass-radius relation is flatter at early times than it was for the linear $S(M)$ function (Fig. \ref{fig:Radius_vs_t0}). Planet contraction however turns to expansion when the planet mass decreases sufficiently. An exception to that is the lowest entropy value, $S_{\rm min} = 7.5$, the pink dashed curve, for which the planet radius always shrinks and eventually becomes equal to the core radius of about $2 R_\oplus$. This is simply the case of a planet that lost all of its gaseous atmosphere and just the core survives. Note that all of the other curves would also arrive at the same final radius but MESA iterations do not always converge when planets expand rapidly.

In the case of a very massive solid core, shown in the bottom panel in Fig. \ref{fig:M_R_SA_with_core}, we observe that the core has a profound effect on the radius of the planet when $M_{\rm p}$ drops below $\sim (2-3)\mj$, which is an order of magnitude higher than $M_{\rm c}$. Planet radius is a monotonically decreasing function of time (decreasing as mass $M_{\rm p}$ decreases) for all but the two highest values of $S_{\rm p}$.

\subsection{A small parameter space study}\label{sec:param_space}

Here we perform a small two-parameter phase study of our model for the planets with internal entropy profile given by eq. \ref{S_heat_exp}. The first parameter is $S_{\rm min}$. The second is the disc feeding rate $\dot M_{\rm feed}$, which is varied from the minimum of $\dot M_{\rm feed} = 5\times 10^{-7}$ \MSunPerYear to the maximum of $4\times 10^{-6}$ \MSunPerYear in seven logarithmically uniform steps. Unlike \S \ref{sec:Bursts_linS}, where the planet was injected in the disc at $a=0.08$~AU, here we inject the planet at $a=0.5$~AU. This is sufficiently far from the TI-unstable region for the disc and the planet to adjust to one another's presence before the planet migrates into the inner disc and EE bursts commence.

\subsubsection{The role of $\dot M_{\rm feed}$.}\label{sec:dotM_feed}

\begin{figure*}
\includegraphics[width=0.99\textwidth]{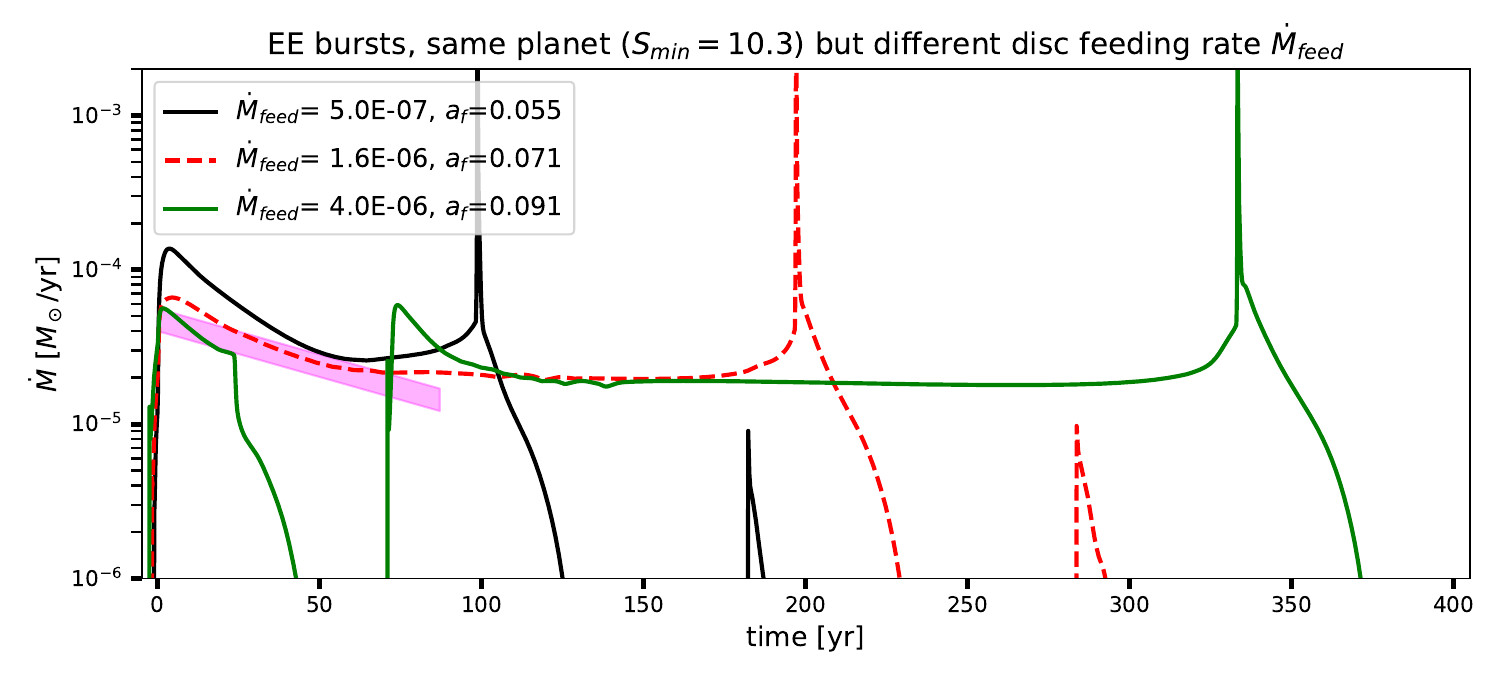}
\caption{FUOR bursts from the planet with identical properties but immersed in discs with different feeding rates at large radius, as marked in the legend. The pink shaded area is eq. \ref{Mdot_fit}. The smaller is $\dot M_{\rm feed}$, the closer to the star the planet needs to be to experience EE-causing conditions, so the bursts are brighter, and their character may change. See \S \ref{sec:dotM_feed}.}
\label{fig:FUOR_Sexp_difMd}
\end{figure*}

Fig. \ref{fig:FUOR_Sexp_difMd} shows stellar mass accretion rates during the 400 years after the beginning of an EE burst for the same planet, $M_{\rm c} = 5\mearth$, $S_{\rm min} = 10.3$, but for three different values of $\dot M_{\rm feed}$ as shown in the legend. We also present the location of where the planet is disrupted, $a_{\rm f}$, in the legend. We shifted the time axis to have $t=0$ in the beginning of the outburst. The pink shaded area is the ``desired" FU Ori $\dot M$ evolution given by eq. \ref{dotM_obs}. 

As expected based on paper I, at a fixed value of $R_{\rm p0}$, the higher is $\dot M_{\rm feed}$, the larger is $a_{\rm f}$, because the disc is hotter and hence the EE condition $R_{\rm p} > R_{\rm B}$ is encountered by the migrating planet earlier. At the same time, planets evaporating closer in to the star produce more powerful outbursts. This  is logical since after the  onset of the outburst the disc around them heats up to higher temperatures than it does for planets evaporated at larger distance. This results in EE bursts that are shorter and brighter at smaller $\dot M_{\rm feed}$ for the same initial $R_{\rm p}$. Similar effects were also seen for planets with $R_{\rm p}$ contracting due to radiative cooling (\S 7.3 in paper I, and fig. 21 in particular).

We note however that this prediction -- brighter bursts in lower $\dot M_{\rm feed}$ discs -- holds here but may not hold in a fully self-consistent calculation. We consider planets of the same $R_{\rm p0}$ and internal structure. In a fully self-consistent calculation where the planets evolve, they contract with time, and are likely to be smaller in discs with smaller $\dot M_{\rm feed}$. TI-EE bursts may disappear completely in older discs with too small $\dot M_{\rm feed}$ (cf. fig. 17 in paper I).

Fig. \ref{fig:FUOR_Sexp_difMd} also shows that the character of EE bursts may change significantly as  $\dot M_{\rm feed}$ varies. There is one continuous EE burst for the two lower values of $\dot M_{\rm feed}$ in the figure, but at $\dot M_{\rm feed} = 4\times 10^{-6}$ \MSunPerYear, an initial burst stutters after only about 20 years because the planet contracts and becomes smaller than $R_{\rm B}$, so EE process terminates. The burst however restarts at $t\approx 70$ years, this time burning the planet till the end.

Looking through the results of our models for different $\dot M_{\rm feed}$ we found that $\dot M_{\rm feed} = 2.2\times 10^{-6}$ \MSunPerYear comes closest to matching both the observed stellar accretion rate on FU Ori and also the location $a \approx (0.07-0.08)$~AU of the suspected planet responsible for the disc hotpot in the model of QPOs observed from FU Ori \cite{PowellEtal12,Siwak21-FUOri-QPOs}. In the next section we explore how planet internal structure affects EE bursts for this particular value of $\dot M_{\rm feed}$.

\subsubsection{The role of planet internal structure}\label{sec:Sexp_bursts}

\begin{figure*}
\includegraphics[width=0.99\textwidth]{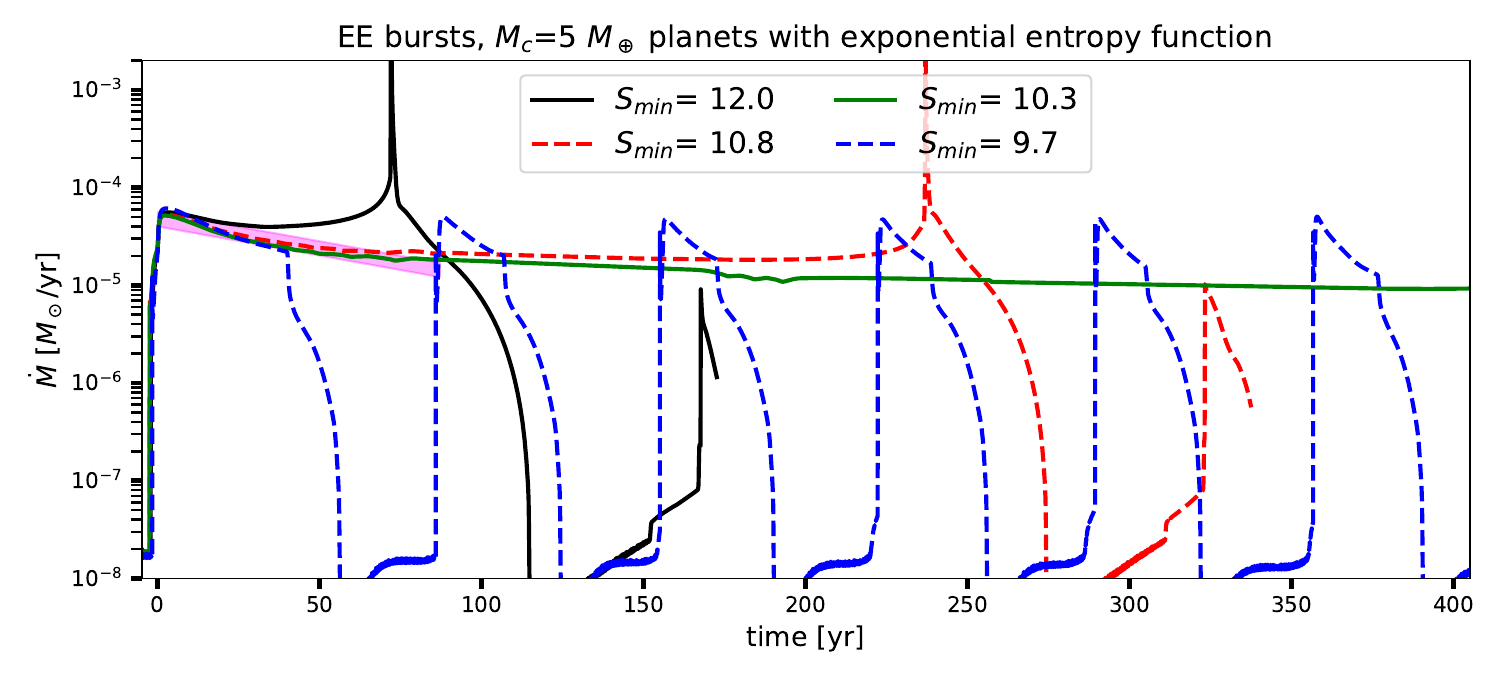}
\includegraphics[width=0.99\textwidth]{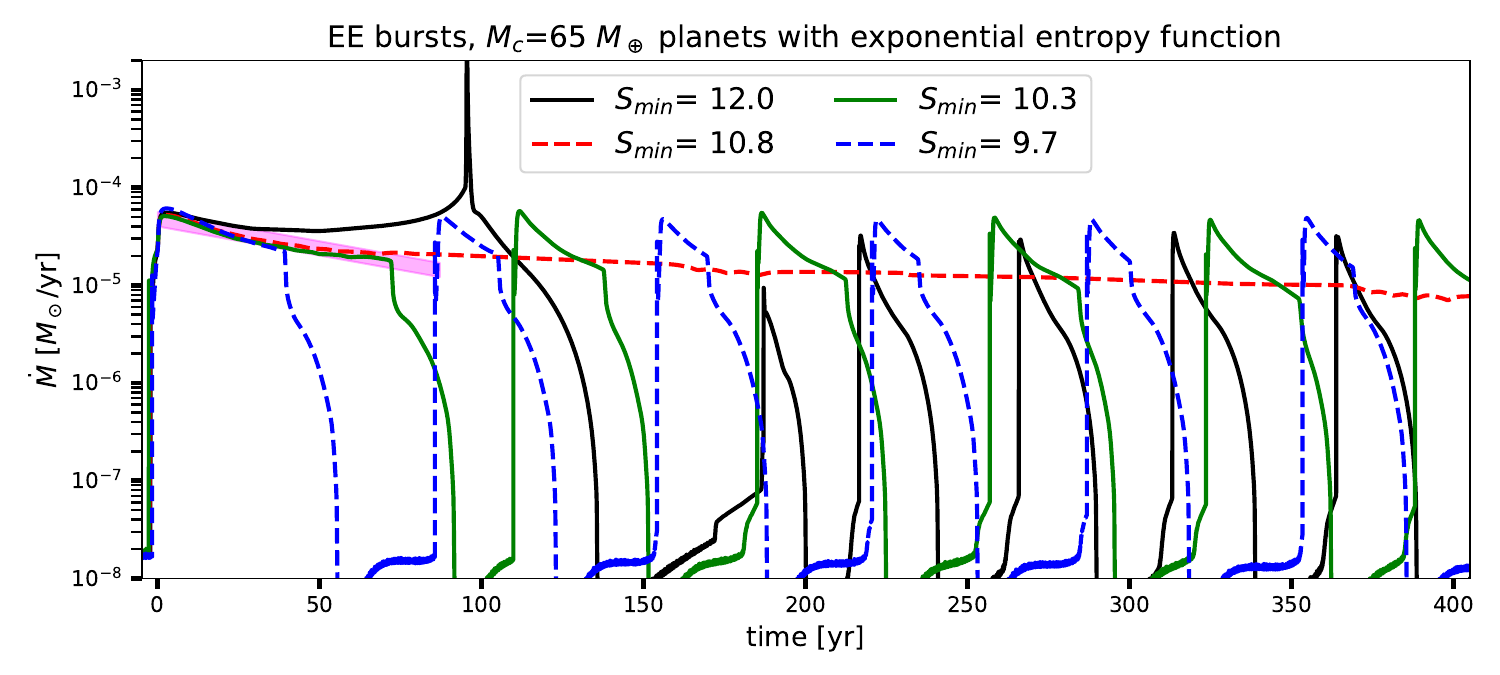}
\caption{Same as Fig. \ref{fig:FUOR_difS} but for planets with exponential entropy profile (\S \ref{sec:M_R_expS}) and solid cores with mass $M_{\rm c} = 5\mearth$ (top panel) and $M_{\rm c} = 65\mearth$ ( bottom panel).}
\label{fig:FUOR_Sexp}
\end{figure*}

Fig. \ref{fig:FUOR_Sexp} shows the resulting stellar accretion rates for $\dot M = 2.2\times 10^{-6}$ \MSunPerYear and various central (minimum) entropy values as given in the legends. The top and bottom panels show the cases $M_{\rm c} = 5\mearth$ and $M_{\rm c} = 65\mearth$, respectively. The planet-star separation at the beginning of the burst is the same, $a_{\rm f} \approx 0.078$~AU, for all of the curves in Fig. \ref{fig:FUOR_Sexp} because the planets are of the same initial radius.

Focusing first on the case of a low mass core, $M_{\rm c} = 5\mearth$, we observe once again that expanding planets (the black curve) lose their mass the quickest, and end up tidally disrupted too soon. The $S_{\rm min} = 10.8$ and 10.3 models produce stellar accretion rates within a factor of two of that needed to explain FU Ori observations, although their $\dot M$ is too flat after $t\sim 70$ years. The smallest value of $S_{\rm min}$ in the top panel of Fig. \ref{fig:FUOR_Sexp} (blue curve) is inconsistent with the observed burst strongly. This planet contracts too fast so its first EE burst terminates quickly. The next TI bursts however reheats the disc sufficiently to restart EE of the planet, albeit for a shorter time. This results in repeated planet-assisted outbursts with mixed characteristics, their nature shifting from EE-dominated in the beginning to practically pure TI later on.

The outbursts for the planet with a more massive core, $M_{\rm c} = 65\mearth$, bottom panel of Fig. \ref{fig:FUOR_Sexp}, are qualitatively similar but there are quantitative differences. Fig. \ref{fig:M_R_SA_with_core} showed that planets tend to have smaller radii if they have a more massive core\footnote{after they lost some mass. Recall that by design our planets have the same initial radius $R_{\rm p}$ at $t<0$.}. This implies that planets with a more massive core are less likely to be tidally disrupted. This is why the outbursts with the highest values of $S_{\rm min}$ in Fig. \ref{fig:FUOR_Sexp} last longer in the bottom panel than they do in the top one. For the same reason the higher $M_{\rm c}$ planets are more likely to fall below the $R_{\rm p} = R_{\rm B}$ line and hence their EE bursts tend to stutter more. For example, the $S_{\rm min} = 10.3$ planet (green curve in the top panel) with a low mass core produces one very long continuous outburst whereas the same value of $S_{\rm min}$ in the bottom panel results in multiple shorter bursts. The simulation Exp-10.3-Mc5 \bref{(the acronym means ``exponential entropy profile with central entropy 10.3 and core mass $5\mearth$)} produces a close although not entirely perfect fit to the FU Ori accretion rate. in \S \ref{sec:SED} we consider this model in greater detail.

\begin{figure*}
\includegraphics[width=0.99\textwidth]{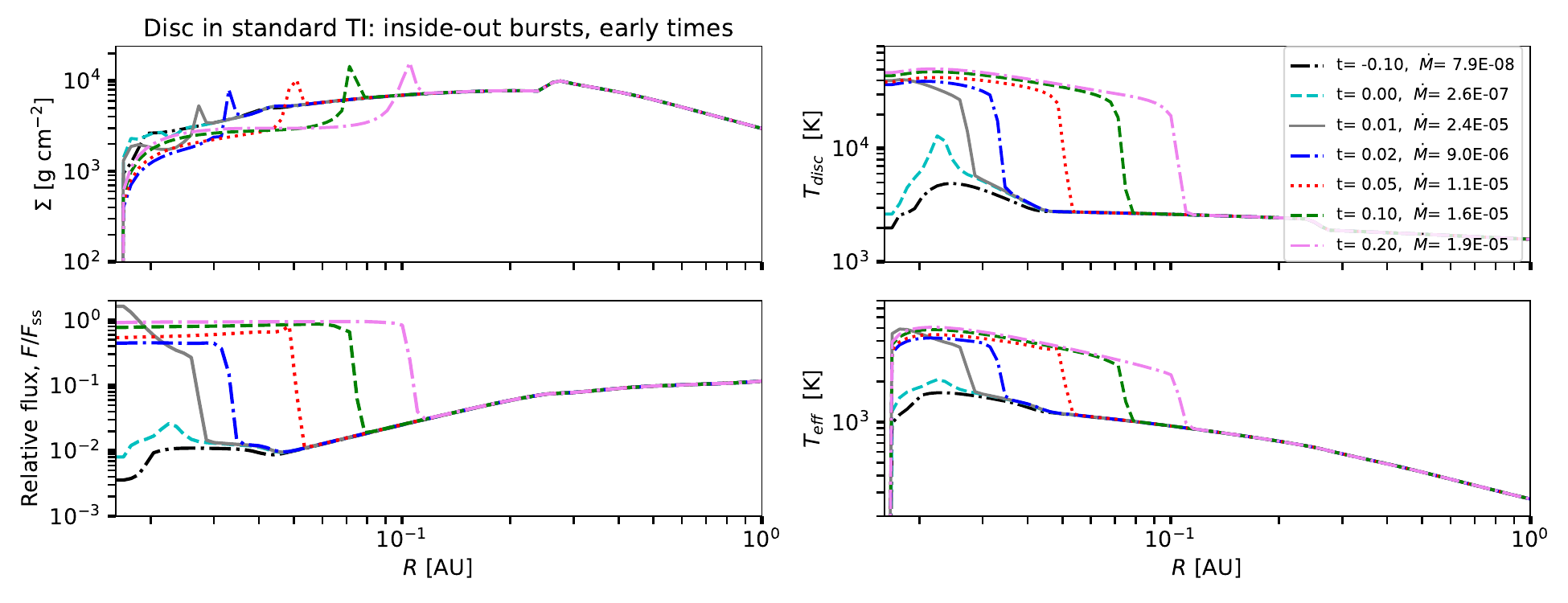}
\includegraphics[width=0.99\textwidth]{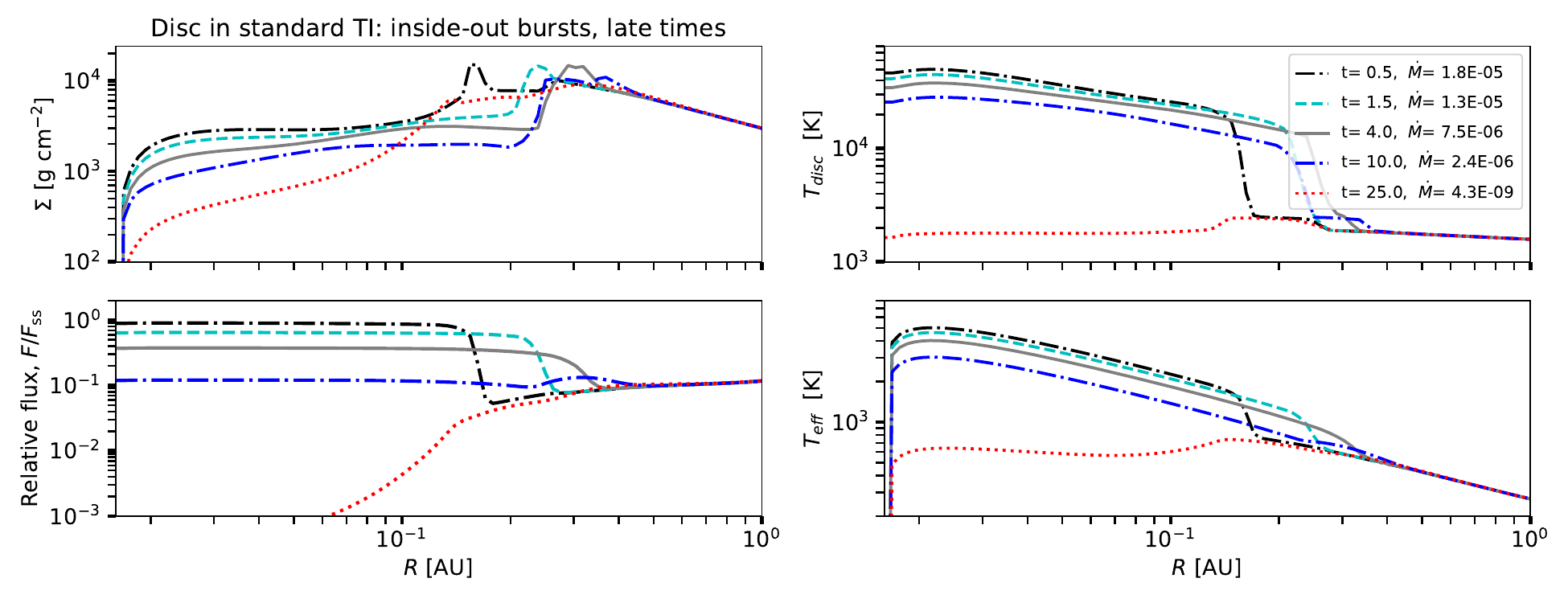}
\caption{Standard planet-free disc structure at early ({\em the top 4 panels}) and late times ({\em the bottom 4 panels}) in the TI outburst shown in Fig. \ref{fig:TI_magnitudes_SED}.}
\label{fig:TI_disc}
\end{figure*}

\begin{figure*}
\includegraphics[width=0.31\textwidth]{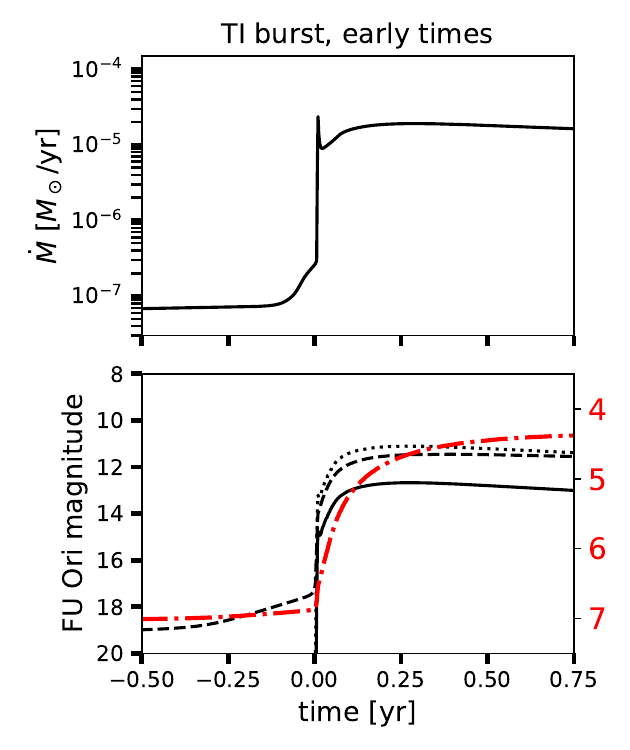}
\includegraphics[width=0.31\textwidth]{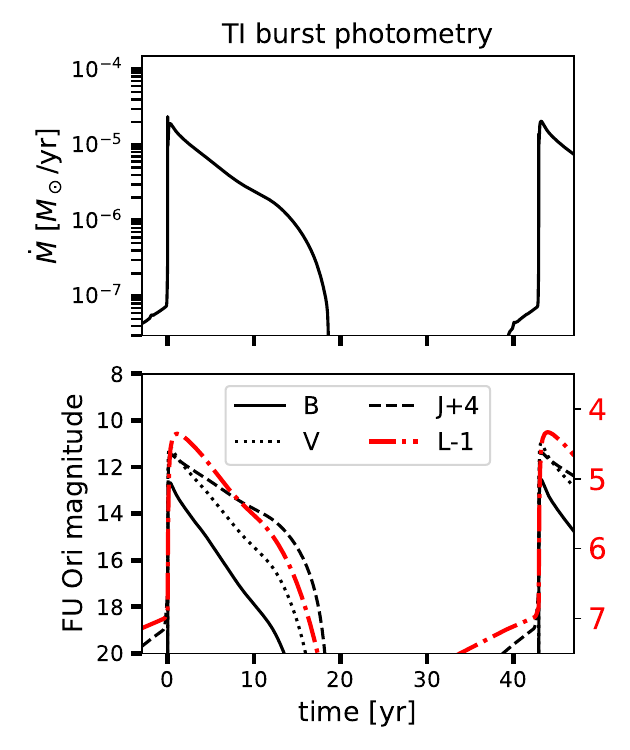}
\includegraphics[width=0.365\textwidth]{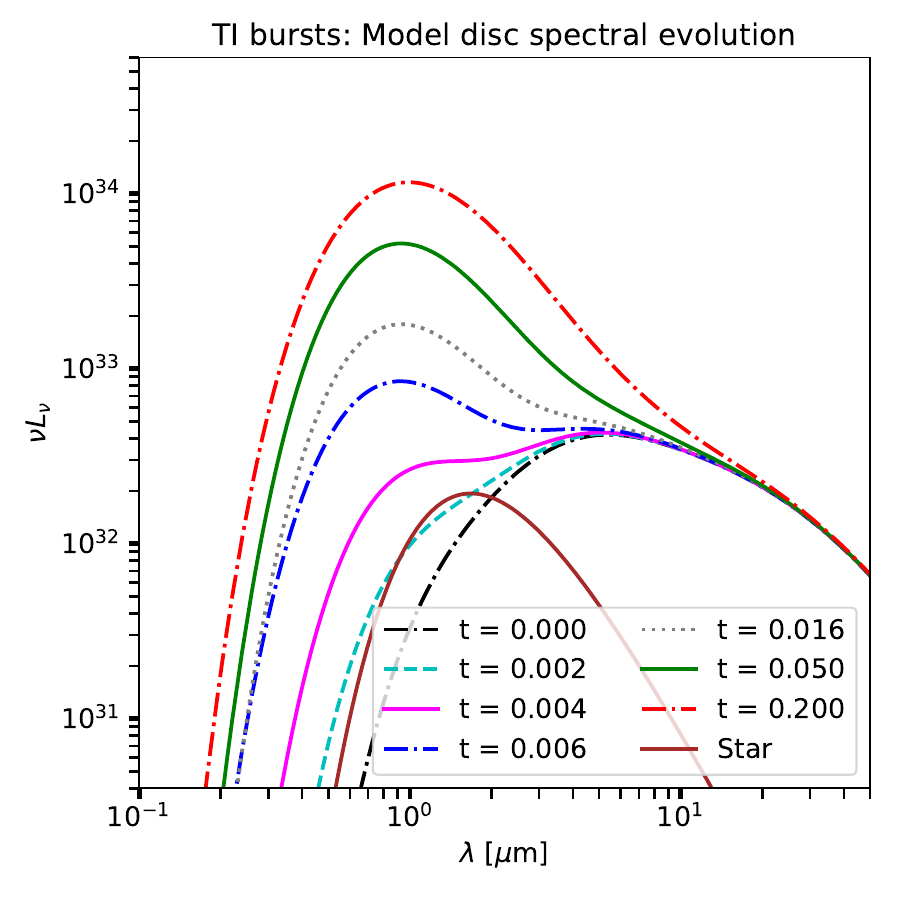}
\caption{Standard planet-free TI outburst observational appearance. {\em Left and Centre panels. Top:} Stellar accretion rate  vs time, {\em bottom:} Magnitude evolution for FU Ori in selected four filters. Note that in the L band the outburst begins weeks to months later than in the other bands, and that the burst lasts a few years longer in J and L filters. {\em Right panel:} SED evolution of the disc during the rise to the maximum light. See \S \ref{sec:standard_TI} for detail.}
\label{fig:TI_magnitudes_SED}
\end{figure*}

\section{Time dependent spectra of unstable discs}\label{sec:SED}

In this section we study how the accretion disc structure and the corresponding Spectral Energy Distribution (SED) vary through the model outbursts for classical TI, TIP-EE, and the DZ-MRI scenarios. The focus is on the early SED evolution as we find that the three models make diverging predictions that can facilitate their observational differentiation.

\subsection{Classical TI bursts: months-long mid-IR delays}\label{sec:standard_TI}

Here we analyse a standard planet-free TI disc behaviour  through one outburst cycle. The disc feeding rate $\dot M_{\rm feed}$ is the same as in Fig. \ref{fig:FUOR_Sexp}, $2.2 \times 10^{-6}$ \MSunPerYear. The resulting stellar accretion rate during the cycle is shown in the centre top panel of Fig. \ref{fig:TI_magnitudes_SED}. The time axis in this section  is shifted to $t=0$ at the moment of the initial very brief stellar accretion rate spike in the outburst (seen in the top left panel of the figure which zooms in on the first year of the burst.)

Fig. \ref{fig:TI_disc} shows disc evolution from $t=-0.1$ to $t=0.2$~yrs in the top 4 panels, whereas the bottom 4 panels show disc evolution from $t=0.5$ until the disc falls deep in quiescence, $t=25$~yrs. Three of the panels show fairly obvious disc characteristics, the disc surface density $\Sigma$, the midplane and effective temperatures, $T_{\rm d}$ and $T_{\rm eff}$, respectively. The relative flux quantity, ${\cal F}$, is the ratio of the \bref{instantaneous} emergent radiation disc flux, $F$, to the radiation flux of a steady-state \cite{Shakura73} disc, $F_{\rm ss}$, for a fixed $\dot M_{\rm ref} = 2\times 10^{-5}$ \MSunPerYear. For a steady-state disc with accretion rate $\dot M$, ${\cal F}=\dot M/\dot M_{\rm ref}=$~const everywhere; deviations from this are tell-tale signs of time-dependency.


The early disc evolution shown in the top 4 panels of Fig. \ref{fig:TI_disc}  reproduces the well known result \citep[e.g.,][]{Bell94,BellEtal95} that classical TI outbursts begin in the inner disc, at a distance of $\sim 2-3$ stellar radii. Time $t=0$ coincides with emergence of an ionisation peak at $R\approx 0.023$~AU. The ionisation fronts propagate inward and outward, reaching the star the quickest. Within just a few days, the accretion rate onto the star rises by two orders of magnitude.  

The two bottom panels in  Fig. \ref{fig:TI_magnitudes_SED} show model magnitudes in the B, V, J and L bands ($\lambda \approx $ 0.545, 0.641, 1.25, 3.5 $\mu m$, respectively). We compute those by assuming that the disc emits local blackbody emission, and integrating this emission over the disc annuli. We assume disc inclination of $i=37^\circ$ and use a constant in time $A_{\rm v}=1.7$ for visual extinction \citep[cf.][]{2022Lykou}. The right panel of Fig. \ref{fig:TI_magnitudes_SED} presents the integrated disc SED evolution during the TI burst rise. We see that the optical (B \& V) and the near-IR J bands rise to the maximum brightness in a matter of weeks whereas the mid-IR L band lags. Quantitatively, a rise of 2 magnitudes in the L band occurs $\sim 2$ months later than the burst onset in the B/V bands. The peak in the L is offset even more, by $\sim 1$ year, with respect to the peaks in the optical bands. This delay is comparable to the ionisation front propagation time through the unstable region. The front propagates outward at the speed $\sim \alpha c_s (H/R) = \alpha (H/R)^2 v_K \sim 0.004 v_K$ \citep{Bell94} for $\alpha=0.1$ and $H/R \sim 0.2$.

The middle panels of Fig. \ref{fig:TI_magnitudes_SED} show that the near-IR and mid-IR emission of the disc outlast the optical burst by a few years. 
The red dotted curve ($t=25$) shows that in quiescence the inner disc essentially disappears. Also note that the disc beyond $R=0.3$~AU varies very little throughout the full outburst cycle; this region of the disc sits stably on the low $\dot M$ neutral Hydrogen solution branch.  


\subsection{TI-EE bursts: months-to-year long optical delays}\label{sec:IR_precursor}

\begin{figure*}
\includegraphics[width=0.99\textwidth]{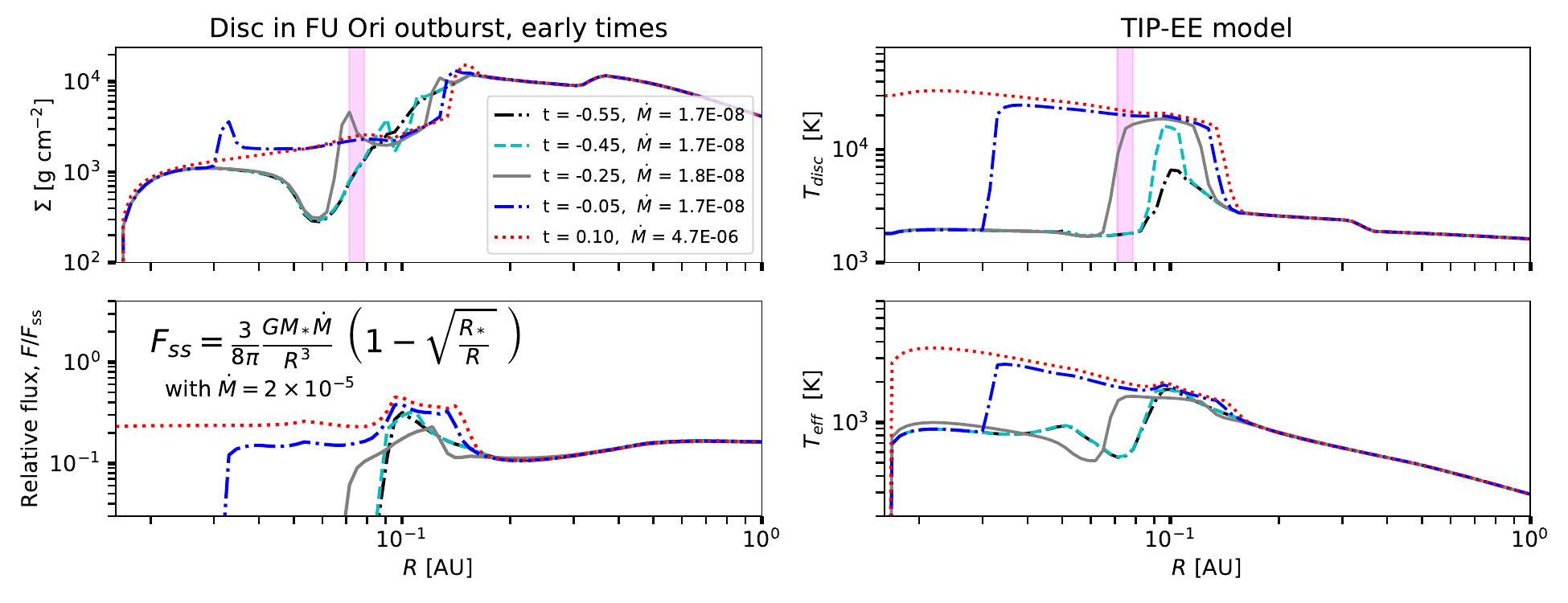}
\includegraphics[width=0.99\textwidth]{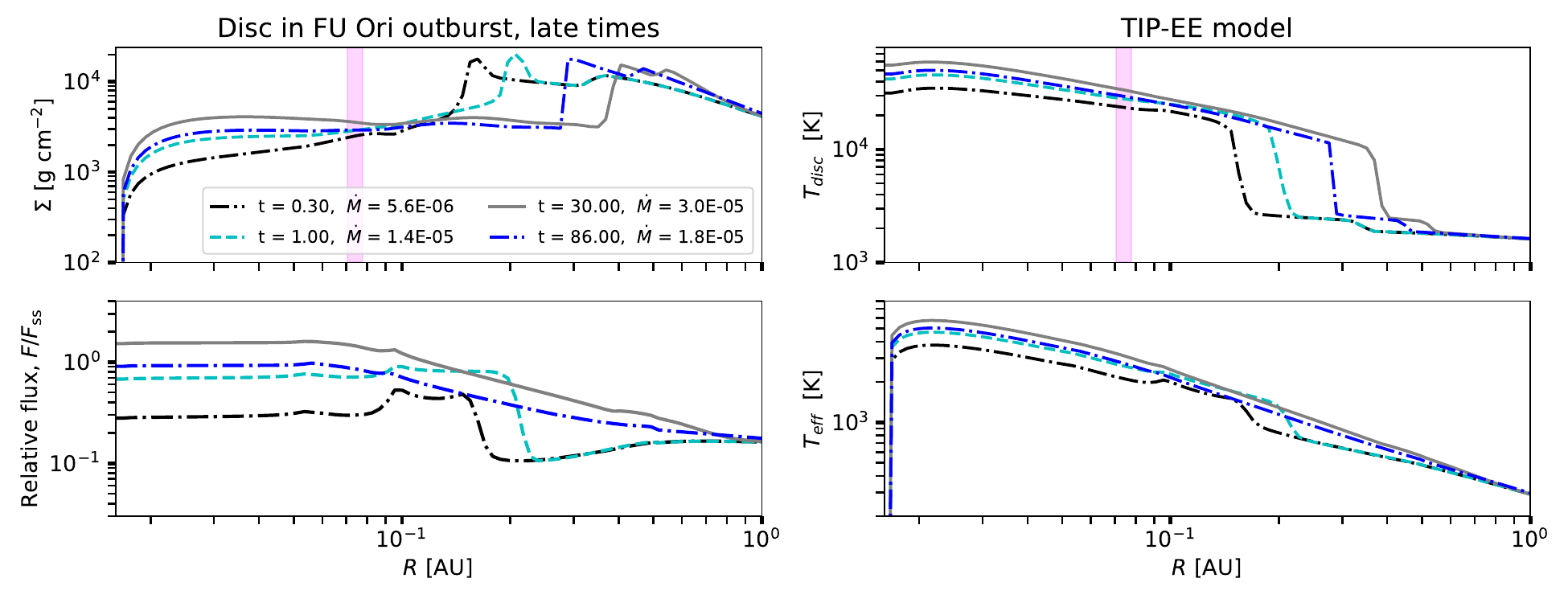}
\caption{Model Exp-10.3-Mc5 disc structure at early ({\em the top 4 panels}) and late ({\em the bottom 4 panels}) times in the outburst.}
\label{fig:FUOR_disc}
\end{figure*}
\begin{figure*}
\includegraphics[width=0.31\textwidth]{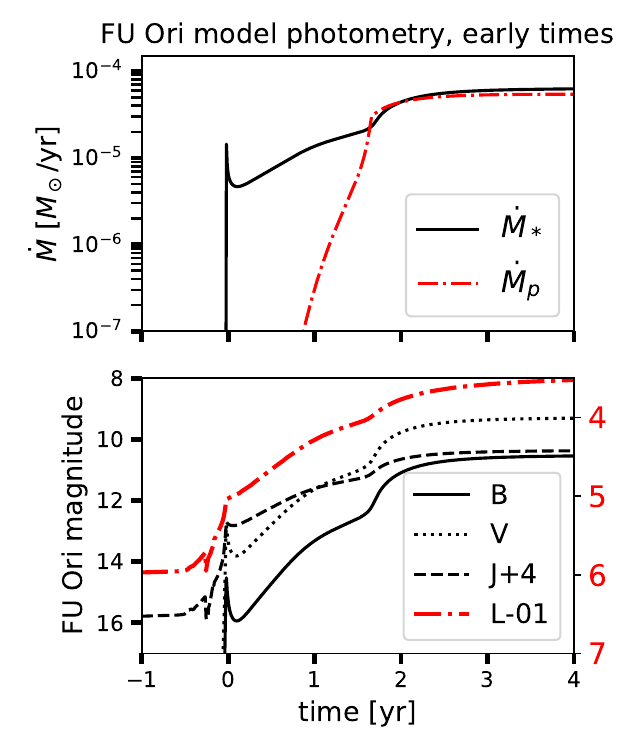}
\includegraphics[width=0.31\textwidth]{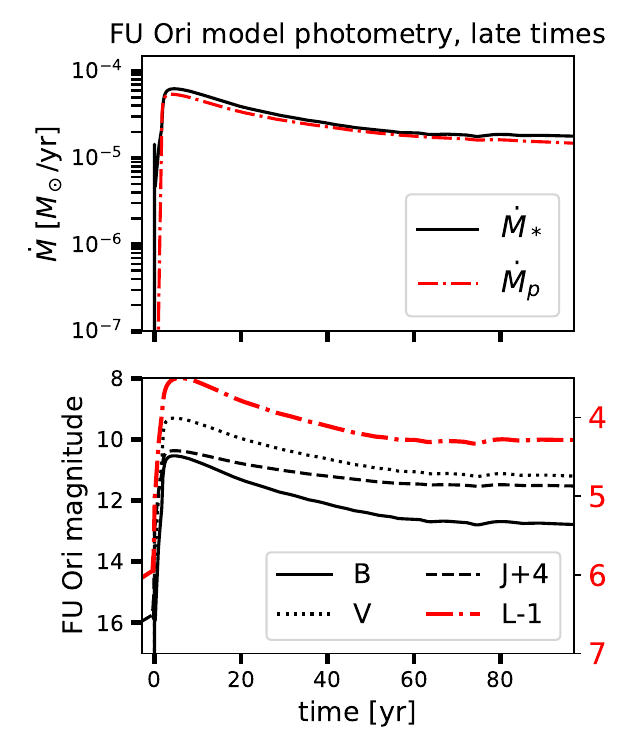}
\includegraphics[width=0.365\textwidth]{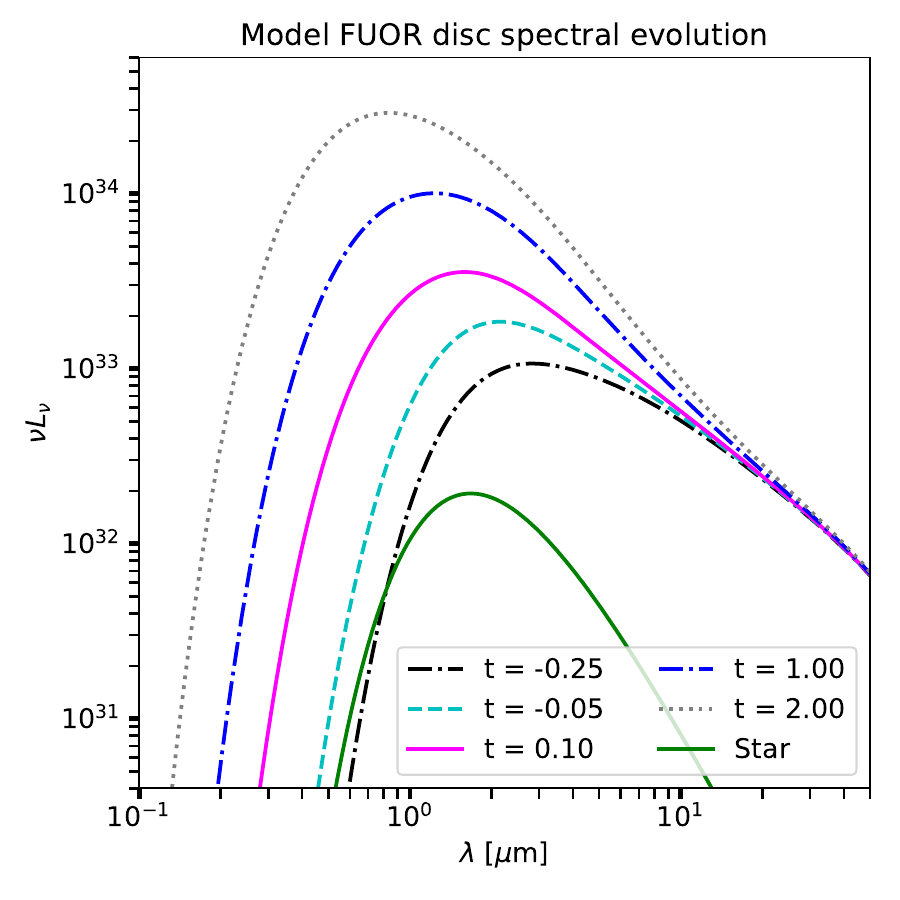}
\caption{{\bf Top:} Stellar accretion rate and planet mass loss rate vs time at early times in model Exp-10.3-Mc5. {\bf Bottom:} Magnitude evolution for FU Ori in selected four filters. Note that in the J and L bands the outburst begins a few months prior to that in the optical.}
\label{fig:Magnitudes_early}
\end{figure*}

Figs. \ref{fig:FUOR_disc} \& \ref{fig:Magnitudes_early} analyse the EE burst of model Exp-10.3-Mc5. 
The vertical violet-shaded band shows the location of the planet (whose orbit barely evolves during the burst). Consider first the early times of the burst, the top 4 panels of Fig. \ref{fig:FUOR_disc}. Prior to the outburst beginning, the gap around the planetary orbit is open. As in the scenario of \cite{LodatoClarke04}, the outburst starts behind the planet, at $R\approx 0.1$~AU. As the disc heats up, $T_{\rm d}$ surges by an order of magnitude. Factoring in the additional change in $\mu$ by about a factor of $4$, the disc $H/R$ at the planet location increases by at least a factor of 5 (cf. fig. 3 in paper I). The value of the Crida parameter $C_{\rm p}$ varies suddenly from sub-unity (gap opened) to a few (see fig. 20 in paper I). The gap is hence filled by the hot gas on the local dynamical time.



Consider the bottom four panels in Fig. \ref{fig:FUOR_disc}. The outburst propagates outward, bringing into the inner disc more matter that was piled up behind the planet, and setting off further heating up of the inner disc. This eventually tips the planet over the $R_{\rm p} > R_{\rm B}$ barrier to EE mass loss. As the planet takes over the role of the main mass supplier to the inner disc, a self-sustained FU Ori outburst begins.

The two left panels in Fig. \ref{fig:Magnitudes_early} show mass fluxes and photometric evolution of the outburst. During the first $\sim 1.5$ years the outburst is powered by the TI-planet \citep{LodatoClarke04} mechanism, with the EE process becoming important only later. Once the planet EE process turns on, the outburst is in the slowly evolving planet sourced-mode for the next $\sim 400$~years (cf. the $S_{\rm min} = 10.3$ curve in Fig. \ref{fig:FUOR_Sexp}). From the bottom left panel and the SED (the right panel of Fig. \ref{fig:Magnitudes_early}) evolution, we see that the outbursts in the NIR band J \& the mid-infrared band L start some $\sim 0.3$~years earlier than in the V or B bands. This is quite different from the classical TI burst behaviour. The outburst rise time to maximum light is $\sim 2$ years in the B and V bands, which is comparable but a little longer than observed \citep[e.g.,][]{Herbig66-FUORi}. Early brightening in mid-IR was observed in several recently discovered FUORs, see \S \ref{sec:Discussion}.


\subsection{DZ-MRI bursts: decade long optical delays}\label{sec:DZ-MRI}


\begin{figure}
\includegraphics[width=1\columnwidth]{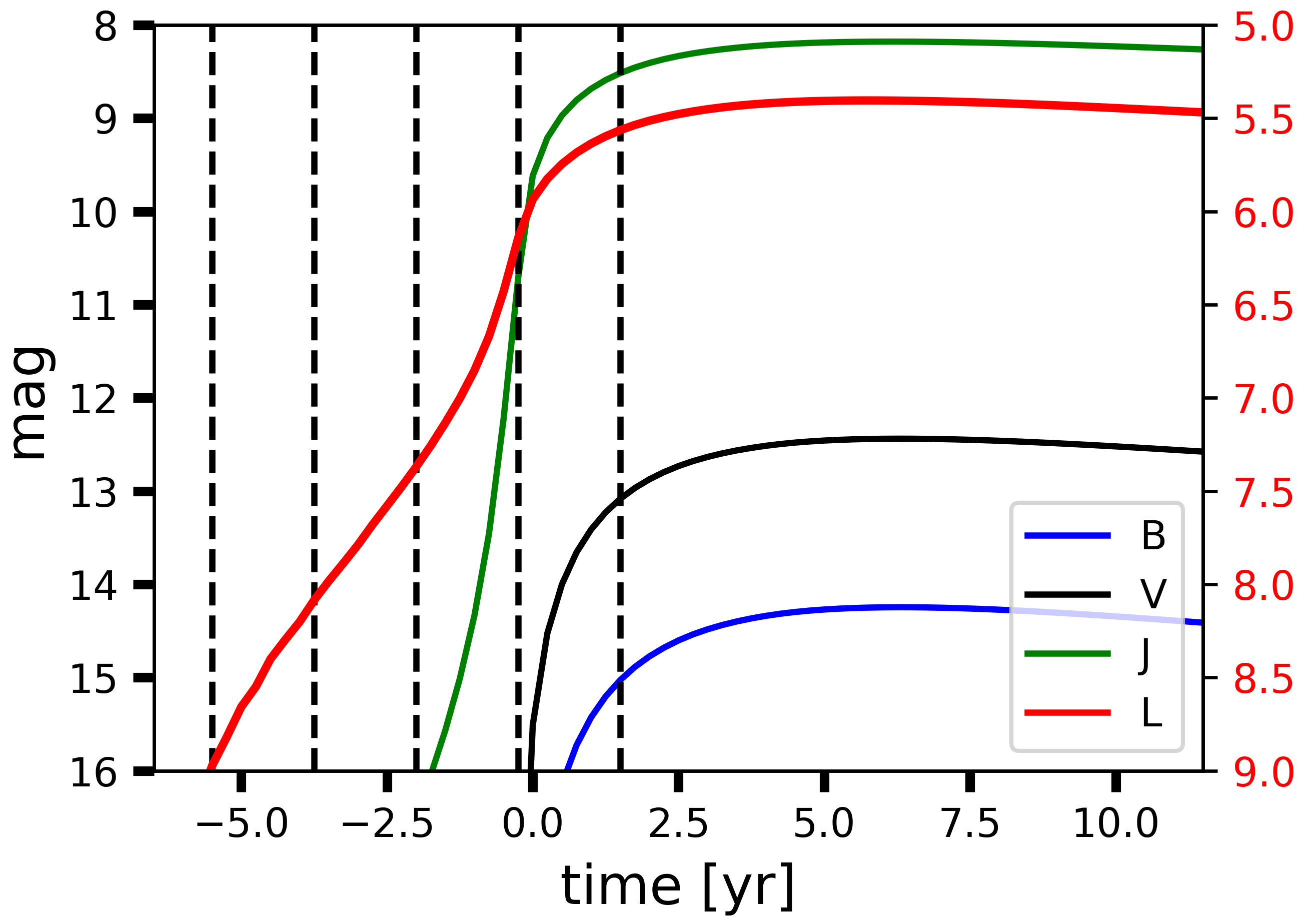}
\caption{Lightcurves in various photometric bands for the $\dot M_{\rm feed} = 8\times 10^{-8}$ \MSunPerYear B23 model shown with the red curve in Fig. \ref{fig:bourdarot_mdot}. Notice that, contrary to the TIP-EE scenario (lower left panel in Fig. \ref{fig:Magnitudes_early}), outbursts in J and L bands precede the optical burst by 2 and 5 years, respectively.}
\label{fig:bourdarot_bands}
\end{figure}

Recently, \cite{Bourdarot-23-FUOR} (B23 hereafter) presented near-IR interferrometric observations of FU Ori and 1D time-dependent disc modelling of the dead zone MRI activation scenario (DZ-MRI) and concluded that the model accounts for the {\em ``outburst region"} size well.
In Figure~\ref{fig:bourdarot_bands} we plot the lightcurves for B23 model with $\dot M_{\rm feed} = 8\times 10^{-8}$ \MSunPerYear in various photometric bands\footnote{Appendix A provides technical detail on how this was computed.}. We note that the B and V band lightcurves in this model are several magnitudes lower than the observed ones. This is because $\dot M$ is a factor of two larger and $R_*$ is about twice smaller in \cite{2022Lykou} than in this model. By adjusting parameters of the model it is possible to match the observed B and V data closer, as shown by B23, however our focus here is on the time lags between the four photometric bands, and that is weakly dependent on $R_*$ or $\dot M$. Figure~\ref{fig:bourdarot_bands} shows that the outburst in J and L bands start approximately two and five years earlier than it does in B and V. 
This is opposite to the behavior seen in classic TI bursts. For the TI-EE scenario the outbursts also start in the mid-IR first, but the delay in the optical emission is months to a year versus at least several years for the DZ-MRI scenario.

\begin{figure*}
\includegraphics[width=0.99\textwidth]{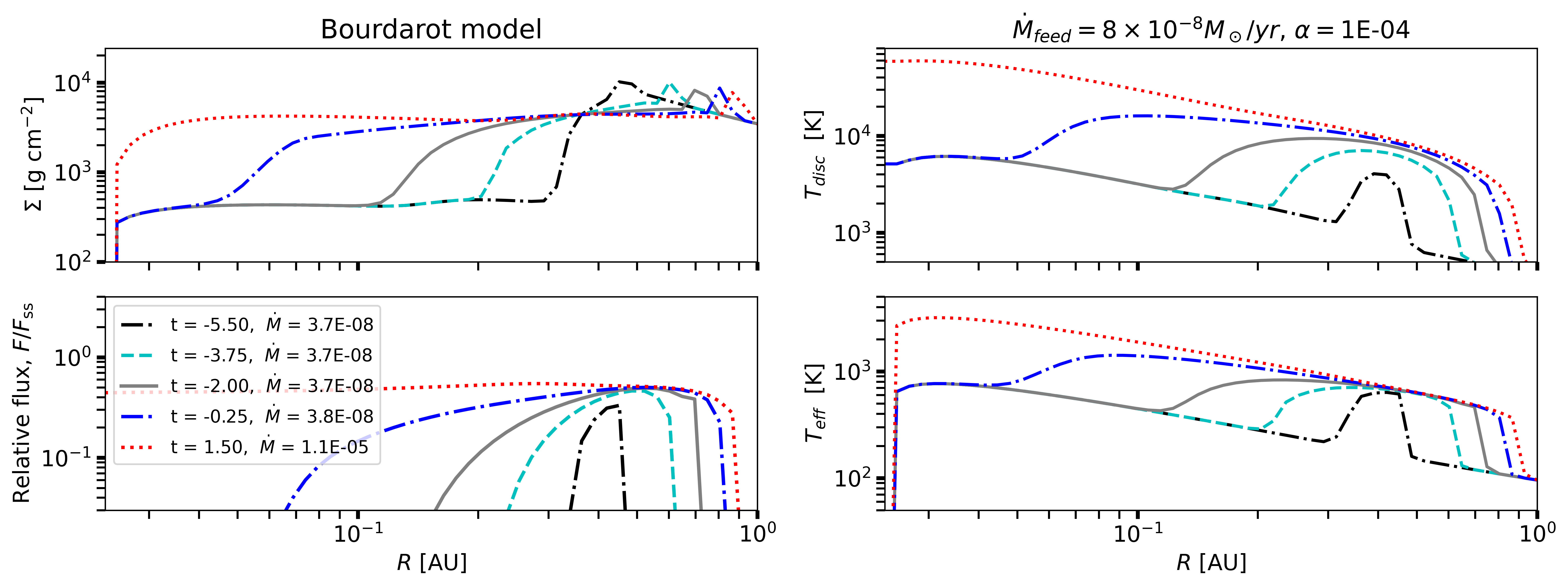}
\caption{Similar to Fig. \ref{fig:FUOR_disc}, but for the $\dot M_{\rm feed} = 8\times 10^{-8}$ \MSunPerYear model shown with the thick red line in Figure~\ref{fig:bourdarot_mdot}. The time corresponds to the time in Figure~\ref{fig:bourdarot_mdot}.}
\label{fig:bourdarot}
\end{figure*}

These results are best understood by considering disc evolution during the MRI activation burst ignition shown in Figure~\ref{fig:bourdarot} for the same ($\dot M_{\rm feed} = 8\times 10^{-8}$ \MSunPerYear B23) DZ-MRI model. The inner edge of MRI inactive zone (the DZ) is at $R\approx0.4$~AU before the outburst. The burst is thermally (rather than GI) triggered when midplane temperature exceeds the critical temperature of 800 K there \citep[interestingly this is similar to the suggestion made by][]{Cleaver-23-FUOR}. As the disc at this location gets ionised, a heating wave propagates both inward and outward. It takes $\sim 5$ years for the ionisation front to reach the star, and this delay is the reason why mid-IR brightens much earlier than do B and V bands. 

Differences in model parameters, such as critical temperature, $T_{\rm crit}$, disc surface density $\Sigma_{\rm crit}$, viscosity parameter values in the DZ and during the burst, $\dot M_{\rm feed}$, will all lead to different outburst $\dot M$ curves. However, such outbursts always start at a significant fraction of an AU from the star \citep[up to 2-3 AU; see][]{ArmitageEtal01,Zhu10-DZ-MRI-1D}. Therefore, the key result of this section -- the mid-IR emission preceding the optical in MRI activation bursts by many years -- is robust to parameter choices. 

When this manuscript was in its final stages, \cite{Cleaver-23-FUOR} presented a detailed study of accretion bursts in DZ-MRI scenario. They assume a constant viscosity parameter, i.e., $\alpha_{\rm hot} = \alpha_{\rm cold} = 0.1$, and they also reduce dust opacity by an order of magnitude to account for dust growth in the disc. Both of these assumptions
modify the location of the critical point $\Sigma_{\rm A}$ where the instability usually sets in on the lower stable branch \citep[see Fig. 2 and Fig. 1 in][respectively]{Bell94,LodatoClarke04}. We believe that these choices straighten the S-curve out and so largely prevent the TI instability from operating, except for relatively small dips in the lightcurves \citep[see figures in][]{Cleaver-23-FUOR}. The authors also investigate a range of pre-burst initial $\Sigma(R)$ profiles and the location of the seed perturbation that sets the MRI instability off, rather than letting the cycles repeat on their own, as we have done here. Despite all these differences, \cite{Cleaver-23-FUOR} also find that for powerful FU Ori type outbursts the location of the starting perturbation is $\sim 1$~AU in the DZ-MRI scenario, and they find that the optical burst lags IR emission by as much as decades. Similar conclusions are obtained in a simpler analytical modelling by \cite{Liu22-FUORs}, see their fig. 21. Thus, irrespective of modelling detail, the optical emission in DZ-MRI bursts lags IR rise by years to several decades.

\subsection{External perturbers}\label{sec:external_perturbers}

\bref{FU Ori is a binary system \citep{Perez20-FUOri}, with the outbursting star FU Ori North referred to simply as FU Ori here. In this paper we completely ignored FU Ori South even though binary interactions \citep{Bonnell92-binary-FUORs} and stellar flybys \citep[e.g.,][]{Vorobyov-21-flyby-FUORs,Borchert-22-flyby-FUORs} were shown to be able to produce powerful accretion outbursts in simulations. However, in application specifically to FU Ori, this scenario is probably not very likely for several reasons: (i) we now know \citep{2022Lykou} that the active disc radius is very small, $\lesssim 0.3$~AU, which is much smaller than predicted by the simulations cited above; (ii) the observed separation of the components in FU Ori exceeds 200 AU. If the two FU Ori components form a bound pair then the orbit must be extremely eccentric for a close passage to affect the inner disc of FU Ori. But this passage would have happened no less than 300 years ago rather than 1937. This scenario is also challenged by the point ``v" below; (iii) If the pair is unbound instead then the stars must be travelling with an uncommonly high relative velocity for typical star cluster environment to yield a close passage in 1937. The probability of a passage of two stars within a few AU or each other in this case is $\sim 10^{-6}$ (privite communication, Andrew Winter). (iv) Accretion outbursts resulting from stellar flybys are strongly peaked events \citep[e.g.,][]{Borchert-22-flyby-FUORs} that do not resemble the surprisingly slow decline of FU Ori's brightness. (v) The two discs around the components of FU Ori are quite smooth \citep{Perez20-FUOri}, not showing any evidence for a recent violent interaction. }

\bref{The binary nature of FU Ori is however important in constraining the large-scale (tens of AU) disc structure and evolution. {\em If} our model for FU Ori is correct then this implies that discs in binary systems can hatch planets via GI and these planets are able to migrate into the inner 0.1 AU by the time the discs look relaxed and GI-spiral free. This will be addressed in future work.}

\section{Discussion}\label{sec:Discussion}



\subsection{Internal structure of GI planets}\label{sec:disc_inside_planets}


In paper I, a power-law mass radius relation for the model planet was assumed. Here, in \S \ref{sec:M_R_linear_entropy} and \S \ref{sec:M_R_expS}, we computed the mass-radius relation of mass losing planets with a stellar evolution code. In such a calculation one specifies an initial condition for the internal structure of the planet, which unfortunately is not well constrained for the youngest GI planets (see \S \ref{sec:toy_expectations}). In realistic planets  composition $Z$ and specific entropy $S$ are functions of $M$, the enclosed mass inside the planet. Here we investigated a simple scenario of a uniform composition gas envelope  (metallicity $Z=0.02$), with a possible presence of an inert solid ($Z=1$) core in the centre, and several forms for $S(M)$. $S(M)$ cannot be a strongly decreasing function of $M$ as vigorous convection ensues, so we  have two cases to consider. ``Standard planets" are those initialised with constant entropy as per default MESA approach \citep{Paxton13-Mesa-2}. Super-adiabatic planets are those with a positive gradient in $S(M)$.
In \S \ref{sec:M_R_linear_entropy} coreless (core mass $M_{\rm c}=0$) planets were studied, whereas in \S \ref{sec:M_R_expS} we allowed for a presence of an inert solid core with $M_{\rm c}>0$. TI-EE outbursts depend strongly on the planet internal structure:

\begin{enumerate}

\item  Standard core-less planets expand rapidly while losing mass. 
Such planets lead to runaway FUOR outbursts whereby planet mass loss keeps increasing (and so stellar $\dot M$) until the planet fills its Hill radius, $R_{\rm H}$, at which point it is disrupted in a powerful and short burst.

\item Super-adiabatic coreless planets with a linear entropy function ($dS/dM = $ const $ > 0$) may first contract and then eventually expand (Fig. \ref{fig:Radius_vs_t0}). For small values of $dS/dM$, the outbursts are similar to those of the standard planets, but tidal disruption of the planet is delayed. For large values of $dS/dM$, the planet contracts so rapidly that Extreme Evaporation process terminates while most of the planet is still intact. This leads to rapidly declining short outbursts. 

\item Century long continuous outbursts such as FU Ori's must be somewhat rare because for this the planet radius must be in a relatively narrow region, the grey area in Fig. \ref{fig:Radius_vs_t0}, between the Bondi radius, $R_{\rm B}$, and $R_{\rm H}$.

\item Bursts powered by EE of coreless planets, $M_{\rm c}=0$, always end in powerful and short tidal disruption bursts.

\item In \S \ref{sec:M_R_expS} we tested an exponential form of entropy function (eq. \ref{S_heat_exp}). The resulting planet $M-R$ relation (Fig. \ref{fig:M_R_SA_with_core}) in general also shows contraction followed by expansion, however solid core presence leads to a rapid contraction when the envelope mass $M-M_{\rm c}\sim M_{\rm c}$. Extreme Evaporation of such planets does not necessarily end in a runaway disruption burst, and there is a planetary remnant with the core of $M_{\rm c}$ and some atmosphere. 

\item Planets that contract while losing mass yield repetitive shorter bursts, such as the green and blue curves in the bottom panel of Fig. \ref{fig:FUOR_difS}. This may explain why FUOR phenomenon is so widespread. \cite{HK96}  estimated that FUOR bursts happen a dozen times in the life of each young star while its growing. Recent observations suggest that these outbursts recur most frequently in class 0 phase \citep{Hsieh19-FU-Ori,Zakri-22-FUOR-Class-0}. For Class II sources,  recurrencce time is $\sim 10^5$ years, an order of magnitude longer than in Class I \citep{Contreras-19-FUOR-statistics}. However since duration of class II phase is longer than class I this may represent a non negligible contribution to FUOR event rate. If each accretion burst required one massive planet then
this would require an extraordinary high rate of planet formation by disc fragmentation, i.e., $\sim$ dozens of GI planets per star. The repetitive bursts we see in Fig. \ref{fig:FUOR_difS} however reduce this requirement significantly, probably making it consistent with a few to ten GI clumps per star formed in disc fragmentation simulations \citep[e.g.,][]{VB06,VB10,ChaNayakshin11a}.

\item The repetitive EE bursts may also be the mode, consistent with the fact that many recently discovered outbursts are not century long events but last for only $\sim$ a dozen years and may repeat \citep[e.g.,][]{Fischer-PPVII}. Indeed, a number of FUORs have now been seen to undergo switching from the burst to near quiescence and back up, e.g., V346 Nor \citep{Kospal-17-V346,Kospal-20-V346}, V899 \citep{Ninan-15-V899,2023Ninan}, V1647 \citep{Ninan-13-V1647}. Such behaviour is not expected for DZ-MRI scenario where periods between bursts are expected to be $\sim (10^3 - 10^4)$ years. For the model parameters explored here, the recurrence time of the repetitive EE bursts is much shorter, e.g., tens of years (Fig. \ref{fig:FUOR_Sexp}), closer to those observed.

\item Our best fit model for FU Ori (the green solid curve in the top panel of Fig. \ref{fig:FUOR_Sexp}) suggests that the planet radius shrank only by $\sim 30$\% from the burst beginning till now, and its mass is $\sim$ half of its initial mass of $6\mj$. 

\end{enumerate}

\subsection{Multi wavelength tests of FUOR outbursts}

Observations of episodic accretion bursts in multiple bands, especially during rapid light curve evolution, is a key tool to test outburst models \citep[as previously suggested by, e.g.,][]{Clarke-90-FUOR,BellEtal95,Bourdarot-23-FUOR,Cleaver-23-FUOR}. In \S \ref{sec:SED} we compared light curve evolution in four bands (optical B, V, and mid-IR J and L) for three models of FUOR bursts: the classical TI, the TI-EE, and the DZ-MRI. 

\begin{enumerate}
    \item As previously found by \cite{BellEtal95}, classical TI bursts start very close to the star and propagate outward (Fig. \ref{fig:TI_disc}). Spectroscopically, such bursts start in the optical, with mid-IR emission rising some months later (Fig. \ref{fig:TI_magnitudes_SED}).
    \item  In our best TI-EE scenario for FU Ori,  the burst starts behind the planet at $R\sim 0.1$~AU, as suggested by \cite{LodatoClarke04}, rather than in the inner disc (Fig. \ref{fig:FUOR_disc}). The burst first becomes apparent in the mid-IR, with the optical emission delayed by a few months (Fig. \ref{fig:Magnitudes_early}) to a year.
    
    \item In the DZ-MRI scenario, the outburst starts at $R\sim (0.5 - 2)$~AU. Such outbursts rise in the IR years to decades before the optical emission does; these delays are much longer than in the TIP-EE model \citep[note that our results reproduce simpler modelling by][cf. their Fig. 21]{Liu22-FUORs}. Further, \cite{Cleaver-23-FUOR} has recently found several decades long optical delays for bright FUORs. 
    
\end{enumerate}

There are only two sources for which this phase of the burst was well observed in the modern era of multiwavelength  observations \citep[see \S 4.1.2 and fig. 6 in][]{Fischer-PPVII}, and in both cases the optical burst started after the IR rises in the lightcurve. In {\em Gaia}-17bpi \citep{Hillenbrand-18-FUOR} the optical delay is at least a year, whereas in {\em Gaia}-18dvy \citep{Szegedi-Elek-20-Gaia-18} the delay is probably somewhat shorter. Classic TI predicts that optical precede IR, so this scenario is ruled out for both sources. DZ-MRI scenario and TIP-EE scenario both predict the right sign for the delay, e.g., the optical coming after the IR. Interestingly, however, \cite{Cleaver-23-FUOR} show that the delays correlate with outburst $\dot M$ in the DZ-MRI picture. In particular, for the weak burst in {\em Gaia}-17bpi, where the peak accretion rate is estimated to be $\sim (1-6) \times 10^{-7}$\MSunPerYear \citep{Rodriguez-22-FUORs}, the delay is predicted to be $\sim 1$ year. For FU Ori like outbursts the delay is several decades.  {\em Gaia}-18dvy peak accretion rate is estimated to be smaller but comparable to FU Ori, e.g., $\sim 6\times 10^{-6}$\MSunPerYear, thus one to two orders of magnitude larger than in {\em Gaia}-17bpi. DZ-MRI scenario hence predict a much longer delay in {\em Gaia}-18dvy compared with {\em Gaia}-17bpi, but this is not observed. While here we studied FU Ori like outbursts and plan to address weaker outbursts in near future, we note that the optical delays cannot be very different from a year in our scenario.  TIP-EE outbursts always start at relatively small radius, $\sim 0.1$~AU. Thus, currently available multiband photometry observations of early lightcurve rises are best consistent with our scenario. We also note that in {\em Gaia}-18dvy the outer radius of the bright (active) disc is surprisingly small, $\sim 0.1$~AU \citep[Section 4.3 in ][]{Szegedi-Elek-20-Gaia-18}. As we showed in paper I, $R_{\rm act}$ values much smaller than $\sim 1$~AU challenge DZ-MRI scenario strongly but are consistent with the TI-EE scenario, especially early on in the outburst.



\section{Aknowledgement}

James Owen is warmly thanked for providing his MESA setup for calculations of planet contraction and evaporation that we modified for our purposes here, and for useful comments on the draft. The authors are grateful to \'Agnes K\'osp\'al for a illuminating discussions of recent FUOR observations. Allona Vazan is thanked for discussions and comments on a very early draft. \bref{Andrew Winter is thanked for discussion of the binary flyby scenario for FU Ori.} The authors acknowledge the funding from the UK Science and Technologies Facilities Council, grant No. ST/S000453/1.  This research used the ALICE High Performance Computing Facility at the University of Leicester, and the DiRAC Data Intensive service at Leicester, operated by the University of Leicester IT Services, which forms part of the STFC DiRAC HPC Facility (www.dirac.ac.uk). For the purpose of open access, the authors have applied a Creative Commons Attribution (CC-BY) licence to any Author Accepted Manuscript version arising.

\section{Data availability}

The data obtained in our simulations can be made available on reasonable request to the corresponding author.

\appendix

\section{DZ-MRI disc modelling}\label{sec:technical}

It has been argued in paper I that MRI activation model for FU Ori \citep[][A01 hereafter]{ArmitageEtal01} is inconsistent with observations of the source for two main reasons: (1) the {\em active disc region} is too large compared with mid-IR interferrometric observations of \cite{2022Lykou} and (2) the inner disc must go through TI cycles which contradicts the surprising long term  stability of FU Ori lightcurve. 

Concerning point (1), after paper I was accepted, \cite{Bourdarot-23-FUOR} (B23 hereafter) showed that their model with MRI activation outburst is in a good agreement with the observations.
This appears to contradict paper I results, however we note the definition differences here. B23 define the {\em outburst} region through its observed brightness, therefore, it is really an emitting region bright in H and K photometric bands. In contrast, \cite{2022Lykou} and in paper I, an {\em active disc} region is defined as the disc region where $\dot M \approx \dot M_*$. Outside of that region the local energy dissipation rate is set to zero in \cite{2022Lykou}; in paper I it is not zero but drops significantly with distance, in qualitative similarity to their model. This is best seen in the bottom left panel of Fig. \ref{fig:FUOR_disc}: the effective energy dissipation rate is a factor of $\sim 5$ smaller in the disc beyond 0.4 AU compared to that in the inner 0.1 AU. 

Our preliminary investigation shows that the size of the {\em emitting region} in the optical or near-IR bands could be independent of what happens in the disc beyond $\sim 0.2$~AU for accretion rates of a few $\times 10^{-5}$\MSunPerYear: the disc at these regions, either active or inactive, emits little. In contrast, mid-IR bands are more sensitive to the disc emission at $\gtrsim 0.1$~AU where  our and MRI activation models may be more divergent. A detailed model comparison must however include irradiation of the outer passive disc by the radiation emitted from the inner disc \citep[with a radiative transfer calculation similar to those performed by, e.g.,][]{Zhu08-FUOri,2022Lykou} and we leave this to future work.

Point (2) appears to be robust, and we see no physically motivated way to turn TI off in the MRI activation scenario. This statement does not contradict previous literature. A01 uses a computational grid of 120 radial mesh points that are uniformly distributed in $R^{1/2}$ from $R \approx0.023$~AU (5$R_{\odot}$) to 40~AU. Such a grid covers the $R<0.1$~AU disc with 4 grid points only and is far coarser than our grid, which is logarithmic in $R$. We found that when we degrade our grid resolution to match A01 inner grid then our disc models also show no TI (whereas TI is present for same model parameters at our default, higher resolution). Similarly, \cite{Zhu09-FUOR-2D-MRI-sims} finds no TI bursts in their 2D simulations of MRI activation scenario when the inner boundary of their computational domain exceeds 0.1 AU, but the instability does appear when the region is resolved in their simulations. For the purposes of this section we cannot follow these approaches since the inner 0.1 AU disc emits almost all of the disc emission in the optical.


B23 explores MRI activation scenario with and without TI included in their 1D time-dependent disc calculation. The no-TI scenario is found to fit FU Ori lightcurves better. To exclude TI, the authors set gas opacity to $\kappa=0.02T^{0.8}$~cm$^2$g$^{-1}$. With this opacity choice, we also find no thermal instability\footnote{This is expected. For thermal instability to take place in a vertically integrated disc model, the condition $d\ln Q^{-}/d\ln T < 1$ needs to be satisfied for a constant $\alpha$ disc, where $Q^{-}$ is the vertically integrated disc cooling rate. This requires \citep{Frank02} disc opacity $\kappa$ to increase with $T$ faster than $T^4$, which is not the case for this shallow power law opacity.} cycles in our disc models \citep[here and in paper I we use the more complete dust and gas opacity from][]{ZhuEtal09}. Although there does not seem to be a clear justification for this simplified opacity form, its use is illuminating as it allows one to contrast outburst spectral evolution of the  MRI activation scenario with that of classical TI and TIP-EE scenarios studied earlier. In the rest of this section we therefore use the ``no-TI" opacity from B23.


\begin{figure*}
\includegraphics[width=2\columnwidth]{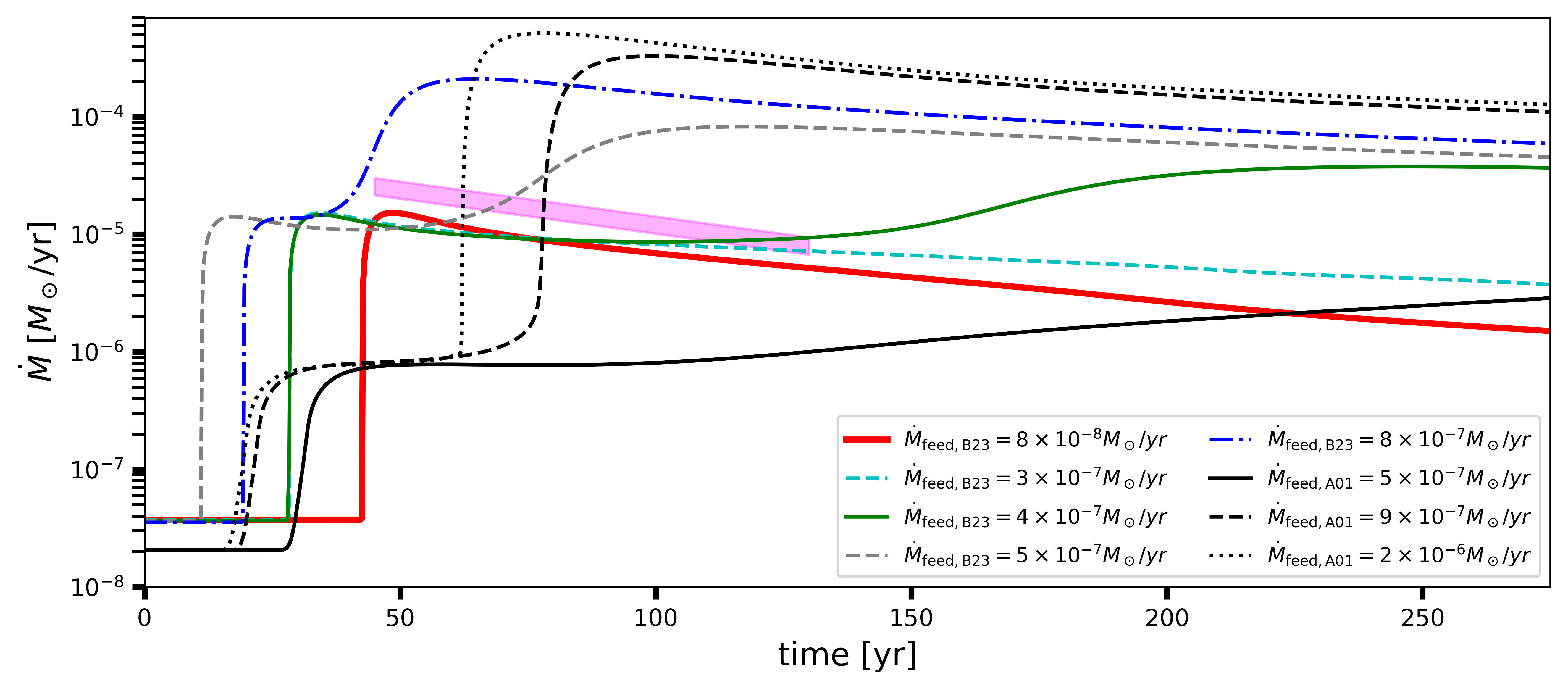}
\caption{Stellar mass accretion rate history for the models similar to B23 models (coloured lines) and A01 models (black lines) with various $\dot M_{\rm feed}$ during a DZ-MRI outburst. The pink shaded area shows the approximate time evolution of observational mass accretion rate for FU Ori during its initial 85 years after the start of outburst. The accretion rate history in model with $\dot M_{\rm feed} = 8\times 10^{-8}$ \MSunPerYear shows the closest agreement with the observational data and is shown with the thick red line.}
\label{fig:bourdarot_mdot}
\end{figure*}

In Figure~\ref{fig:bourdarot_mdot} we show the time evolution of mass accretion rate onto the star for five models that follow disc viscosity choices as in  B23 (coloured lines) for  various values of $\dot M_{\rm feed}$. We also present three models with parameter choices exactly as in A01 (black lines)\footnote{The two models are very similar, however have some differences in parameter values. The outer radius of the disc and the stellar mass are 40~AU and $1\msun$ in the model from A01, while in the model from B23 they are 30~AU and $0.5\msun$, respectively. Furthermore, in B23 a modified prescription for the $\alpha$ parameter in MRI active zone is used (see their Eq.~(4)). Also, in B23 the authors are using $\Sigma_{\rm crit}=10$~g~cm$^-2$, which is an order of magnitude lower than in A01. }. All the curves were shifted in time to display them on the same scale.

In the B23 models, the initial sharp rise in mass accretion rate reaches $\dot M_{\rm feed} \approx 5\times 10^{-5}$ \MSunPerYear during about 1 year. Depending on the value of $\dot M_{\rm feed}$, stellar $\dot M$ shows either a second rise with a gradual decrease or a gradual decrease. We believe that the two-step rise to maximum in some of our models is due to our more complete equation of state that includes the change in the mean molecular weight of the gas as Hydrogen is ionised, while this was fixed at $2.3 m_p$ by B23 and A01. Such a two-step rise in brightness was not observed in FU Ori, but for what follows it will not be important anyway.


As in Fig. \ref{fig:FUOR_Sexp}, the pink shaded area in Fig. \ref{fig:bourdarot_mdot} is the approximate time evolution of  mass accretion rate for FU Ori according to \cite{2022Lykou}. 
The accretion rates in B23 models with $\dot M_{\rm feed} < 4\times 10^{-7}$ \MSunPerYear have rise times similar to the observed ones for FU Ori, and the model with $\dot M_{\rm feed} = 8\times 10^{-8}$ \MSunPerYear (the thick red line) comes closest to the desired $\dot M$ evolution.

\section{TI presence in MRI activation scenario}\label{sec:TI_appendix}

\begin{figure}
\includegraphics[width=1\columnwidth]{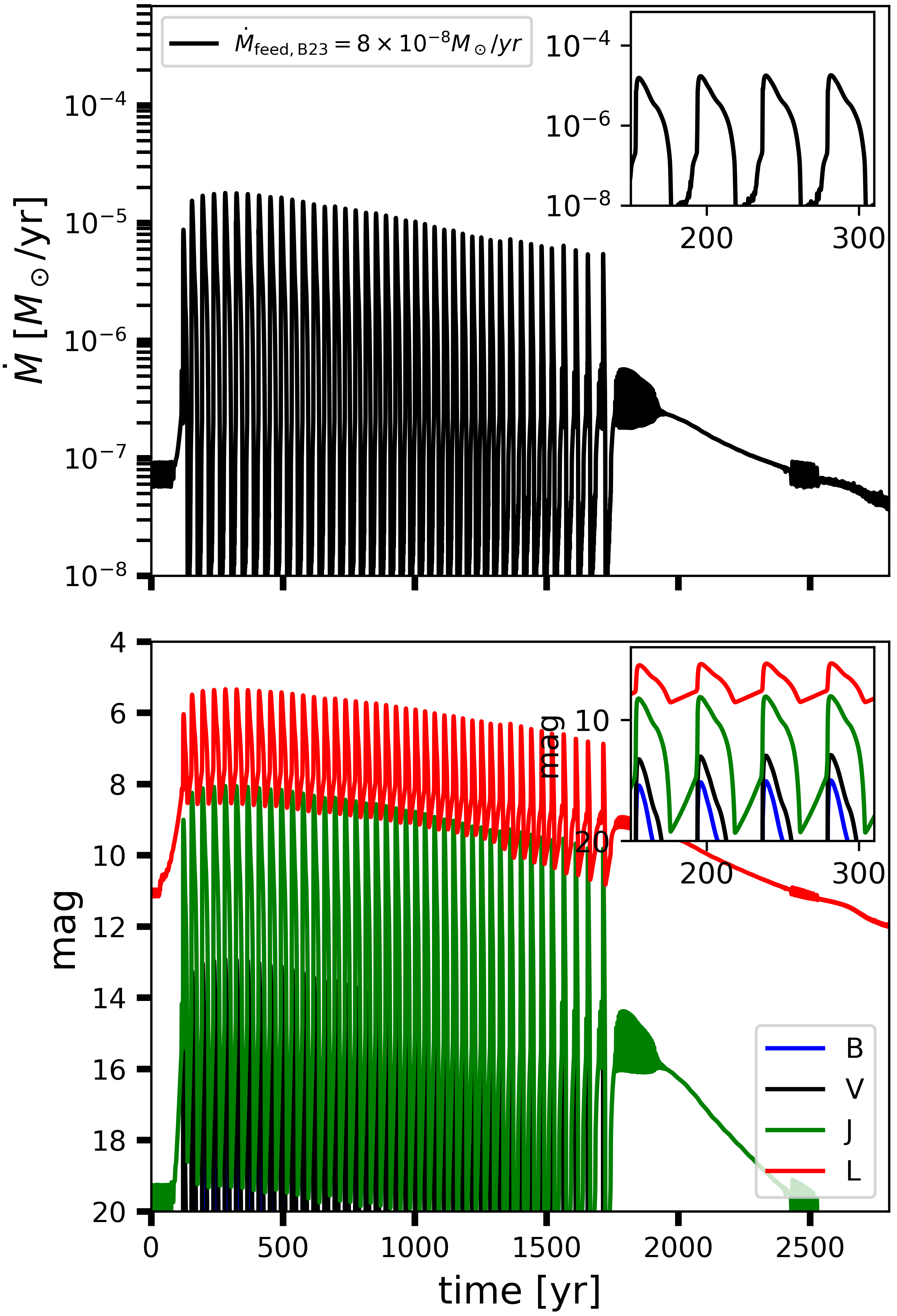}
\caption{Similar to Fig. \ref{fig:bourdarot_mdot} but with realistic disc opacities.}
\label{fig:bourdarot_mdot_TI}
\end{figure}

We claimed in paper I that DZ-MRI scenario model for FU Ori should show TI outbursts and this challenges this scenario. In \S \ref{sec:DZ-MRI} we used simplified opacity form from B23 to study bursts which would start at the inner edge of the DZ. Here we run the same model but with realistic opacities from \citet{ZhuEtal09}. We find that TI is developing in the inner disc for all the $\dot M_{\rm feed}$ values shown in Fig. \ref{fig:bourdarot_mdot}. In Figure~\ref{fig:bourdarot_mdot_TI} we show the stellar mass accretion rates (top panel) and the lightcurves in various bands (bottom panel) zoomed-in on one of the MRI bursts in the model with $\dot M_{\rm feed} = 8\times 10^{-8}$ \MSunPerYear. The duration of DZ-MRI burst is about 2500~yrs, but during the outburst much shorter (a few decades long) TI outbursts are present. In the inset we show a sequence of such TI outbursts. Clearly, such a variable accretion is not consistent with FU Ori observations.

\bibliographystyle{mnras}
\bibliography{nayakshin}

\begin{thebibliography}{}
\makeatletter
\relax
\def\mn@urlcharsother{\let\do\@makeother \do\$\do\&\do\#\do\^\do\_\do\%\do\~}
\def\mn@doi{\begingroup\mn@urlcharsother \@ifnextchar [ {\mn@doi@}
  {\mn@doi@[]}}
\def\mn@doi@[#1]#2{\def\@tempa{#1}\ifx\@tempa\@empty \href
  {http://dx.doi.org/#2} {doi:#2}\else \href {http://dx.doi.org/#2} {#1}\fi
  \endgroup}
\def\mn@eprint#1#2{\mn@eprint@#1:#2::\@nil}
\def\mn@eprint@arXiv#1{\href {http://arxiv.org/abs/#1} {{\tt arXiv:#1}}}
\def\mn@eprint@dblp#1{\href {http://dblp.uni-trier.de/rec/bibtex/#1.xml}
  {dblp:#1}}
\def\mn@eprint@#1:#2:#3:#4\@nil{\def\@tempa {#1}\def\@tempb {#2}\def\@tempc
  {#3}\ifx \@tempc \@empty \let \@tempc \@tempb \let \@tempb \@tempa \fi \ifx
  \@tempb \@empty \def\@tempb {arXiv}\fi \@ifundefined
  {mn@eprint@\@tempb}{\@tempb:\@tempc}{\expandafter \expandafter \csname
  mn@eprint@\@tempb\endcsname \expandafter{\@tempc}}}

\bibitem[\protect\citeauthoryear{{Armitage}}{{Armitage}}{2015}]{Armitage15-review}
{Armitage} P.~J.,  2015, arXiv e-prints, \href
  {https://ui.adsabs.harvard.edu/abs/2015arXiv150906382A} {p. arXiv:1509.06382}

\bibitem[\protect\citeauthoryear{{Armitage}, {Livio}  \& {Pringle}}{{Armitage}
  et~al.}{2001}]{ArmitageEtal01}
{Armitage} P.~J.,  {Livio} M.,   {Pringle} J.~E.,  2001, \mn@doi [\mnras]
  {10.1046/j.1365-8711.2001.04356.x}, \href
  {http://adsabs.harvard.edu/abs/2001MNRAS.324..705A} {324, 705}

\bibitem[\protect\citeauthoryear{{Audard} et~al.,}{{Audard}
  et~al.}{2014}]{AudardEtal14}
{Audard} M.,  et~al., 2014, \mn@doi [Protostars and Planets VI]
  {10.2458/azu_uapress_9780816531240-ch017}, \href
  {http://adsabs.harvard.edu/abs/2014prpl.conf..387A} {pp 387--410}

\bibitem[\protect\citeauthoryear{{Baehr} \& {Klahr}}{{Baehr} \&
  {Klahr}}{2019}]{Baehr19-pebble-accretion}
{Baehr} H.,  {Klahr} H.,  2019, \mn@doi [\apj] {10.3847/1538-4357/ab2f85},
  \href {https://ui.adsabs.harvard.edu/abs/2019ApJ...881..162B} {881, 162}

\bibitem[\protect\citeauthoryear{{Bate}, {Tricco}  \& {Price}}{{Bate}
  et~al.}{2014}]{Bate-14-2nd-core}
{Bate} M.~R.,  {Tricco} T.~S.,   {Price} D.~J.,  2014, \mn@doi [\mnras]
  {10.1093/mnras/stt1865}, \href
  {https://ui.adsabs.harvard.edu/abs/2014MNRAS.437...77B} {437, 77}

\bibitem[\protect\citeauthoryear{{Bell} \& {Lin}}{{Bell} \&
  {Lin}}{1994}]{Bell94}
{Bell} K.~R.,  {Lin} D.~N.~C.,  1994, \mn@doi [\apj] {10.1086/174206}, \href
  {http://ukads.nottingham.ac.uk/cgi-bin/nph-bib_query?bibcode=1994ApJ...427..987B&db_key=AST}
  {427, 987}

\bibitem[\protect\citeauthoryear{{Bell}, {Lin}, {Hartmann}  \& {Kenyon}}{{Bell}
  et~al.}{1995}]{BellEtal95}
{Bell} K.~R.,  {Lin} D.~N.~C.,  {Hartmann} L.~W.,   {Kenyon} S.~J.,  1995,
  \mn@doi [\apj] {10.1086/175612}, \href
  {http://adsabs.harvard.edu/abs/1995ApJ...444..376B} {444, 376}

\bibitem[\protect\citeauthoryear{{Bhandare}, {Kuiper}, {Henning}, {Fendt},
  {Flock}  \& {Marleau}}{{Bhandare} et~al.}{2020}]{2Bhandare-20-2nd-cores}
{Bhandare} A.,  {Kuiper} R.,  {Henning} T.,  {Fendt} C.,  {Flock} M.,
  {Marleau} G.-D.,  2020, \mn@doi [\aap] {10.1051/0004-6361/201937029}, \href
  {https://ui.adsabs.harvard.edu/abs/2020A&A...638A..86B} {638, A86}

\bibitem[\protect\citeauthoryear{{Bodenheimer}}{{Bodenheimer}}{1974}]{Bodenheimer74}
{Bodenheimer} P.,  1974, \mn@doi [Icarus] {10.1016/0019-1035(74)90050-5}, \href
  {http://ukads.nottingham.ac.uk/abs/1974Icar...23..319B} {23, 319}

\bibitem[\protect\citeauthoryear{{Bodenheimer}, {Grossman}, {Decampli}, {Marcy}
   \& {Pollack}}{{Bodenheimer} et~al.}{1980}]{BodenheimerEtal80}
{Bodenheimer} P.,  {Grossman} A.~S.,  {Decampli} W.~M.,  {Marcy} G.,
  {Pollack} J.~B.,  1980, \mn@doi [\icarus] {10.1016/0019-1035(80)90012-3},
  \href {http://adsabs.harvard.edu/abs/1980Icar...41..293B} {41, 293}

\bibitem[\protect\citeauthoryear{{Bodenheimer}, {Laughlin}  \&
  {Lin}}{{Bodenheimer} et~al.}{2003}]{Bodenheimer-03-Tidal-Heating}
{Bodenheimer} P.,  {Laughlin} G.,   {Lin} D. N.~C.,  2003, \mn@doi [\apj]
  {10.1086/375565}, \href
  {https://ui.adsabs.harvard.edu/abs/2003ApJ...592..555B} {592, 555}

\bibitem[\protect\citeauthoryear{{Boley} \& {Durisen}}{{Boley} \&
  {Durisen}}{2010}]{BoleyDurisen10}
{Boley} A.~C.,  {Durisen} R.~H.,  2010, \mn@doi [\apj]
  {10.1088/0004-637X/724/1/618}, \href
  {http://adsabs.harvard.edu/abs/2010ApJ...724..618B} {724, 618}

\bibitem[\protect\citeauthoryear{{Boley}, {Hayfield}, {Mayer}  \&
  {Durisen}}{{Boley} et~al.}{2010}]{BoleyEtal10}
{Boley} A.~C.,  {Hayfield} T.,  {Mayer} L.,   {Durisen} R.~H.,  2010, \mn@doi
  [Icarus] {10.1016/j.icarus.2010.01.015}, \href
  {http://ukads.nottingham.ac.uk/abs/2010Icar..207..509B} {207, 509}

\bibitem[\protect\citeauthoryear{{Bonnell} \& {Bastien}}{{Bonnell} \&
  {Bastien}}{1992}]{Bonnell92-binary-FUORs}
{Bonnell} I.,  {Bastien} P.,  1992, \mn@doi [\apjl] {10.1086/186663}, \href
  {https://ui.adsabs.harvard.edu/abs/1992ApJ...401L..31B} {401, L31}

\bibitem[\protect\citeauthoryear{{Borchert}, {Price}, {Pinte}  \&
  {Cuello}}{{Borchert} et~al.}{2022}]{Borchert-22-flyby-FUORs}
{Borchert} E. M.~A.,  {Price} D.~J.,  {Pinte} C.,   {Cuello} N.,  2022, \mn@doi
  [\mnras] {10.1093/mnrasl/slab123}, \href
  {https://ui.adsabs.harvard.edu/abs/2022MNRAS.510L..37B} {510, L37}

\bibitem[\protect\citeauthoryear{{Boss}}{{Boss}}{1998}]{Boss98}
{Boss} A.~P.,  1998, \mn@doi [\apj] {10.1086/306036}, \href
  {http://adsabs.harvard.edu/abs/1998ApJ...503..923B} {503, 923}

\bibitem[\protect\citeauthoryear{{Bourdarot} et~al.,}{{Bourdarot}
  et~al.}{2023}]{Bourdarot-23-FUOR}
{Bourdarot} G.,  et~al., 2023, \mn@doi [arXiv e-prints]
  {10.48550/arXiv.2304.13414}, \href
  {https://ui.adsabs.harvard.edu/abs/2023arXiv230413414B} {p. arXiv:2304.13414}

\bibitem[\protect\citeauthoryear{{Burrows} et~al.,}{{Burrows}
  et~al.}{1997}]{BurrowsEtal97}
{Burrows} A.,  et~al., 1997, \apj, \href
  {http://adsabs.harvard.edu/abs/1997ApJ...491..856B} {491, 856}

\bibitem[\protect\citeauthoryear{{Cha} \& {Nayakshin}}{{Cha} \&
  {Nayakshin}}{2011}]{ChaNayakshin11a}
{Cha} S.-H.,  {Nayakshin} S.,  2011, \mn@doi [\mnras]
  {10.1111/j.1365-2966.2011.18953.x}, \href
  {http://adsabs.harvard.edu/abs/2011MNRAS.415.3319C} {415, 3319}

\bibitem[\protect\citeauthoryear{{Clarke}, {Lin}  \& {Pringle}}{{Clarke}
  et~al.}{1990}]{Clarke-90-FUOR}
{Clarke} C.~J.,  {Lin} D.~N.~C.,   {Pringle} J.~E.,  1990, \mn@doi [\mnras]
  {10.1093/mnras/242.3.439}, \href
  {https://ui.adsabs.harvard.edu/abs/1990MNRAS.242..439C} {242, 439}

\bibitem[\protect\citeauthoryear{{Cleaver}, {Hartmann}  \& {Bae}}{{Cleaver}
  et~al.}{2023}]{Cleaver-23-FUOR}
{Cleaver} J.,  {Hartmann} L.,   {Bae} J.,  2023, \mn@doi [\mnras]
  {10.1093/mnras/stad1784}, \href
  {https://ui.adsabs.harvard.edu/abs/2023MNRAS.tmp.1701C} {}

\bibitem[\protect\citeauthoryear{{Contreras Pe{\~n}a}, {Naylor}  \&
  {Morrell}}{{Contreras Pe{\~n}a} et~al.}{2019}]{Contreras-19-FUOR-statistics}
{Contreras Pe{\~n}a} C.,  {Naylor} T.,   {Morrell} S.,  2019, \mn@doi [\mnras]
  {10.1093/mnras/stz1019}, \href
  {https://ui.adsabs.harvard.edu/abs/2019MNRAS.486.4590C} {486, 4590}

\bibitem[\protect\citeauthoryear{{Fischer}, {Hillenbrand}, {Herczeg},
  {Johnstone}, {K{\'o}sp{\'a}l}  \& {Dunham}}{{Fischer}
  et~al.}{2022}]{Fischer-PPVII}
{Fischer} W.~J.,  {Hillenbrand} L.~A.,  {Herczeg} G.~J.,  {Johnstone} D.,
  {K{\'o}sp{\'a}l} {\'A}.,   {Dunham} M.~M.,  2022, arXiv e-prints, \href
  {https://ui.adsabs.harvard.edu/abs/2022arXiv220311257F} {p. arXiv:2203.11257}

\bibitem[\protect\citeauthoryear{{Frank}, {King}  \& {Raine}}{{Frank}
  et~al.}{2002}]{Frank02}
{Frank} J.,  {King} A.,   {Raine} D.~J.,  2002, {Accretion Power in
  Astrophysics: Third Edition}.
Accretion Power in Astrophysics, by Juhan Frank and Andrew King and Derek
  Raine, pp.~398.~ISBN 0521620538.~Cambridge, UK: Cambridge University Press,
  February 2002.

\bibitem[\protect\citeauthoryear{{Freedman}, {Marley}  \& {Lodders}}{{Freedman}
  et~al.}{2008}]{Freedman08-opacity}
{Freedman} R.~S.,  {Marley} M.~S.,   {Lodders} K.,  2008, \mn@doi [\apjs]
  {10.1086/521793}, \href
  {https://ui.adsabs.harvard.edu/abs/2008ApJS..174..504F} {174, 504}

\bibitem[\protect\citeauthoryear{{Ginzburg} \& {Sari}}{{Ginzburg} \&
  {Sari}}{2015}]{GinzburgSari15-inflated-hotJ}
{Ginzburg} S.,  {Sari} R.,  2015, \mn@doi [\apj] {10.1088/0004-637X/803/2/111},
  \href {https://ui.adsabs.harvard.edu/abs/2015ApJ...803..111G} {803, 111}

\bibitem[\protect\citeauthoryear{{Graboske}, {Olness}, {Pollack}  \&
  {Grossman}}{{Graboske} et~al.}{1975}]{GraboskeEtal75}
{Graboske} Jr. H.~C.,  {Olness} R.~J.,  {Pollack} J.~B.,   {Grossman} A.~S.,
  1975, \mn@doi [\apj] {10.1086/153689}, \href
  {http://adsabs.harvard.edu/abs/1975ApJ...199..265G} {199, 265}

\bibitem[\protect\citeauthoryear{{Hameury}}{{Hameury}}{2020}]{Hameury-20-review}
{Hameury} J.~M.,  2020, \mn@doi [Advances in Space Research]
  {10.1016/j.asr.2019.10.022}, \href
  {https://ui.adsabs.harvard.edu/abs/2020AdSpR..66.1004H} {66, 1004}

\bibitem[\protect\citeauthoryear{{Hartmann} \& {Kenyon}}{{Hartmann} \&
  {Kenyon}}{1985}]{HartmannK85-FUORs}
{Hartmann} L.,  {Kenyon} S.~J.,  1985, \mn@doi [\apj] {10.1086/163713}, \href
  {https://ui.adsabs.harvard.edu/abs/1985ApJ...299..462H} {299, 462}

\bibitem[\protect\citeauthoryear{{Hartmann} \& {Kenyon}}{{Hartmann} \&
  {Kenyon}}{1996}]{HK96}
{Hartmann} L.,  {Kenyon} S.~J.,  1996, \mn@doi [\araa]
  {10.1146/annurev.astro.34.1.207}, \href
  {http://adsabs.harvard.edu/abs/1996ARA%26A..34..207H} {34, 207}

\bibitem[\protect\citeauthoryear{{Helled} \& {Schubert}}{{Helled} \&
  {Schubert}}{2008}]{HS08}
{Helled} R.,  {Schubert} G.,  2008, \mn@doi [Icarus]
  {10.1016/j.icarus.2008.08.002}, \href
  {http://ukads.nottingham.ac.uk/abs/2008Icar..198..156H} {198, 156}

\bibitem[\protect\citeauthoryear{{Helled}, {Podolak}  \& {Kovetz}}{{Helled}
  et~al.}{2008}]{HelledEtal08}
{Helled} R.,  {Podolak} M.,   {Kovetz} A.,  2008, \mn@doi [Icarus]
  {10.1016/j.icarus.2008.01.007}, \href
  {http://ukads.nottingham.ac.uk/abs/2008Icar..195..863H} {195, 863}

\bibitem[\protect\citeauthoryear{{Herbig}}{{Herbig}}{1966}]{Herbig66-FUORi}
{Herbig} G.~H.,  1966, \mn@doi [Vistas in Astronomy]
  {10.1016/0083-6656(66)90025-0}, \href
  {https://ui.adsabs.harvard.edu/abs/1966VA......8..109H} {8, 109}

\bibitem[\protect\citeauthoryear{{Herbig}, {Petrov}  \& {Duemmler}}{{Herbig}
  et~al.}{2003}]{Herbig03-FUOR}
{Herbig} G.~H.,  {Petrov} P.~P.,   {Duemmler} R.,  2003, \mn@doi [\apj]
  {10.1086/377194}, \href
  {https://ui.adsabs.harvard.edu/abs/2003ApJ...595..384H} {595, 384}

\bibitem[\protect\citeauthoryear{{Hillenbrand} et~al.,}{{Hillenbrand}
  et~al.}{2018}]{Hillenbrand-18-FUOR}
{Hillenbrand} L.~A.,  et~al., 2018, \mn@doi [\apj] {10.3847/1538-4357/aaf414},
  \href {https://ui.adsabs.harvard.edu/abs/2018ApJ...869..146H} {869, 146}

\bibitem[\protect\citeauthoryear{{Hirose}}{{Hirose}}{2015}]{Hirose15}
{Hirose} S.,  2015, \mn@doi [\mnras] {10.1093/mnras/stv203}, \href
  {https://ui.adsabs.harvard.edu/abs/2015MNRAS.448.3105H} {448, 3105}

\bibitem[\protect\citeauthoryear{{Hsieh}, {Murillo}, {Belloche}, {Hirano},
  {Walsh}, {van Dishoeck}, {J{\o}rgensen}  \& {Lai}}{{Hsieh}
  et~al.}{2019}]{Hsieh19-FU-Ori}
{Hsieh} T.-H.,  {Murillo} N.~M.,  {Belloche} A.,  {Hirano} N.,  {Walsh} C.,
  {van Dishoeck} E.~F.,  {J{\o}rgensen} J.~K.,   {Lai} S.-P.,  2019, \mn@doi
  [\apj] {10.3847/1538-4357/ab425a}, \href
  {https://ui.adsabs.harvard.edu/abs/2019ApJ...884..149H} {884, 149}

\bibitem[\protect\citeauthoryear{{Humphries} \& {Nayakshin}}{{Humphries} \&
  {Nayakshin}}{2018}]{HN18}
{Humphries} R.~J.,  {Nayakshin} S.,  2018, \mn@doi [\mnras]
  {10.1093/mnras/sty569}, \href
  {https://ui.adsabs.harvard.edu/abs/2018MNRAS.477..593H} {477, 593}

\bibitem[\protect\citeauthoryear{{Humphries} \& {Nayakshin}}{{Humphries} \&
  {Nayakshin}}{2019}]{HN19-planet-disruption}
{Humphries} J.,  {Nayakshin} S.,  2019, \mn@doi [\mnras]
  {10.1093/mnras/stz2497}, \href
  {https://ui.adsabs.harvard.edu/abs/2019MNRAS.489.5187H} {489, 5187}

\bibitem[\protect\citeauthoryear{{Jackson}, {Greenberg}  \& {Barnes}}{{Jackson}
  et~al.}{2008}]{Jackson08-Tidal-Heaing}
{Jackson} B.,  {Greenberg} R.,   {Barnes} R.,  2008, \mn@doi [\apj]
  {10.1086/587641}, \href
  {https://ui.adsabs.harvard.edu/abs/2008ApJ...681.1631J} {681, 1631}

\bibitem[\protect\citeauthoryear{{Kenyon}, {Kolotilov}, {Ibragimov}  \&
  {Mattei}}{{Kenyon} et~al.}{2000}]{KenyonK-2000-FUOR}
{Kenyon} S.~J.,  {Kolotilov} E.~A.,  {Ibragimov} M.~A.,   {Mattei} J.~A.,
  2000, \mn@doi [\apj] {10.1086/308515}, \href
  {https://ui.adsabs.harvard.edu/abs/2000ApJ...531.1028K} {531, 1028}

\bibitem[\protect\citeauthoryear{{Kippenhahn}, {Weigert}  \&
  {Weiss}}{{Kippenhahn} et~al.}{2013}]{Kippenhahn13-book}
{Kippenhahn} R.,  {Weigert} A.,   {Weiss} A.,  2013, {Stellar Structure and
  Evolution}, \mn@doi{10.1007/978-3-642-30304-3.
}

\bibitem[\protect\citeauthoryear{{K{\'o}sp{\'a}l}, {{\'A}brah{\'a}m},
  {Westhues}  \& {Haas}}{{K{\'o}sp{\'a}l} et~al.}{2017}]{Kospal-17-V346}
{K{\'o}sp{\'a}l} {\'A}.,  {{\'A}brah{\'a}m} P.,  {Westhues} C.,   {Haas} M.,
  2017, \mn@doi [\aap] {10.1051/0004-6361/201629447}, \href
  {https://ui.adsabs.harvard.edu/abs/2017A&A...597L..10K} {597, L10}

\bibitem[\protect\citeauthoryear{{K{\'o}sp{\'a}l}, {Szab{\'o}},
  {{\'A}brah{\'a}m}, {Kraus}, {Takami}, {Lucas}, {Contreras Pe{\~n}a}  \&
  {Udalski}}{{K{\'o}sp{\'a}l} et~al.}{2020}]{Kospal-20-V346}
{K{\'o}sp{\'a}l} {\'A}.,  {Szab{\'o}} Z.~M.,  {{\'A}brah{\'a}m} P.,  {Kraus}
  S.,  {Takami} M.,  {Lucas} P.~W.,  {Contreras Pe{\~n}a} C.,   {Udalski} A.,
  2020, \mn@doi [\apj] {10.3847/1538-4357/ab6174}, \href
  {https://ui.adsabs.harvard.edu/abs/2020ApJ...889..148K} {889, 148}

\bibitem[\protect\citeauthoryear{{Kuiper}, {Klahr}, {Beuther}  \&
  {Henning}}{{Kuiper} et~al.}{2010}]{Kuiper10-RHD}
{Kuiper} R.,  {Klahr} H.,  {Beuther} H.,   {Henning} T.,  2010, \mn@doi [\apj]
  {10.1088/0004-637X/722/2/1556}, \href
  {https://ui.adsabs.harvard.edu/abs/2010ApJ...722.1556K} {722, 1556}

\bibitem[\protect\citeauthoryear{{Larson}}{{Larson}}{1969}]{Larson69}
{Larson} R.~B.,  1969, \mnras, \href
  {http://ukads.nottingham.ac.uk/cgi-bin/nph-bib_query?bibcode=1969MNRAS.145..271L&db_key=AST}
  {145, 271}

\bibitem[\protect\citeauthoryear{{Lasota}}{{Lasota}}{2001}]{Lasota01-Review}
{Lasota} J.-P.,  2001, \mn@doi [\nar] {10.1016/S1387-6473(01)00112-9}, \href
  {https://ui.adsabs.harvard.edu/abs/2001NewAR..45..449L} {45, 449}

\bibitem[\protect\citeauthoryear{{Liu} et~al.,}{{Liu}
  et~al.}{2022}]{Liu22-FUORs}
{Liu} H.,  et~al., 2022, \mn@doi [\apj] {10.3847/1538-4357/ac84d2}, \href
  {https://ui.adsabs.harvard.edu/abs/2022ApJ...936..152L} {936, 152}

\bibitem[\protect\citeauthoryear{{Lodato} \& {Clarke}}{{Lodato} \&
  {Clarke}}{2004}]{LodatoClarke04}
{Lodato} G.,  {Clarke} C.~J.,  2004, \mn@doi [\mnras]
  {10.1111/j.1365-2966.2004.08112.x}, \href
  {http://adsabs.harvard.edu/abs/2004MNRAS.353..841L} {353, 841}

\bibitem[\protect\citeauthoryear{{Lykou} et~al.,}{{Lykou}
  et~al.}{2022}]{2022Lykou}
{Lykou} F.,  et~al., 2022, \mn@doi [\aap] {10.1051/0004-6361/202142788}, \href
  {https://ui.adsabs.harvard.edu/abs/2022A&A...663A..86L} {663, A86}

\bibitem[\protect\citeauthoryear{{Masunaga} \& {Inutsuka}}{{Masunaga} \&
  {Inutsuka}}{2000}]{Masunaga00}
{Masunaga} H.,  {Inutsuka} S.-i.,  2000, \mn@doi [\apj] {10.1086/308439}, \href
  {http://ukads.nottingham.ac.uk/abs/2000ApJ...531..350M} {531, 350}

\bibitem[\protect\citeauthoryear{{McCrea} \& {Williams}}{{McCrea} \&
  {Williams}}{1965}]{McCreaWilliams65}
{McCrea} W.~H.,  {Williams} I.~P.,  1965, Royal Society of London Proceedings
  Series A, \href {http://adsabs.harvard.edu/abs/1965RSPSA.287..143M} {287,
  143}

\bibitem[\protect\citeauthoryear{{Nayakshin}}{{Nayakshin}}{2010}]{Nayakshin10a}
{Nayakshin} S.,  2010, \mn@doi [\mnras] {10.1111/j.1365-2966.2010.17289.x},
  \href {http://adsabs.harvard.edu/abs/2010MNRAS.408.2381N} {408, 2381}

\bibitem[\protect\citeauthoryear{{Nayakshin}}{{Nayakshin}}{2011}]{Nayakshin10b}
{Nayakshin} S.,  2011, \mn@doi [\mnras] {10.1111/j.1365-2966.2011.18230.x},
  \href {http://adsabs.harvard.edu/abs/2011MNRAS.413.1462N} {413, 1462}

\bibitem[\protect\citeauthoryear{{Nayakshin} \& {Lodato}}{{Nayakshin} \&
  {Lodato}}{2012}]{NayakshinLodato12}
{Nayakshin} S.,  {Lodato} G.,  2012, \mn@doi [\mnras]
  {10.1111/j.1365-2966.2012.21612.x}, \href
  {http://adsabs.harvard.edu/abs/2012MNRAS.426...70N} {426, 70}

\bibitem[\protect\citeauthoryear{{Nayakshin}, {Elbakyan}  \&
  {Rosotti}}{{Nayakshin} et~al.}{2022}]{Nayakshin22-ALMA-CA}
{Nayakshin} S.,  {Elbakyan} V.,   {Rosotti} G.,  2022, \mn@doi [\mnras]
  {10.1093/mnras/stac833}, \href
  {https://ui.adsabs.harvard.edu/abs/2022MNRAS.512.6038N} {512, 6038}

\bibitem[\protect\citeauthoryear{{Nayakshin}, {Owen}  \&
  {Elbakyan}}{{Nayakshin} et~al.}{2023}]{Nayakshin-23-FUOR}
{Nayakshin} S.,  {Owen} J.~E.,   {Elbakyan} V.,  2023, \mn@doi [\mnras]
  {10.1093/mnras/stad1392}, \href
  {https://ui.adsabs.harvard.edu/abs/2023MNRAS.tmp.1409N} {}

\bibitem[\protect\citeauthoryear{{Ninan}, {Ojha}, {Bhatt}, {Ghosh}, {Mohan},
  {Mallick}, {Tamura}  \& {Henning}}{{Ninan} et~al.}{2013}]{Ninan-13-V1647}
{Ninan} J.~P.,  {Ojha} D.~K.,  {Bhatt} B.~C.,  {Ghosh} S.~K.,  {Mohan} V.,
  {Mallick} K.~K.,  {Tamura} M.,   {Henning} T.,  2013, \mn@doi [\apj]
  {10.1088/0004-637X/778/2/116}, \href
  {https://ui.adsabs.harvard.edu/abs/2013ApJ...778..116N} {778, 116}

\bibitem[\protect\citeauthoryear{{Ninan} et~al.,}{{Ninan}
  et~al.}{2015}]{Ninan-15-V899}
{Ninan} J.~P.,  et~al., 2015, \mn@doi [\apj] {10.1088/0004-637X/815/1/4}, \href
  {https://ui.adsabs.harvard.edu/abs/2015ApJ...815....4N} {815, 4}

\bibitem[\protect\citeauthoryear{{Ninan}, {Mahadevan}  \& {HPF Team}}{{Ninan}
  et~al.}{2023}]{2023Ninan}
{Ninan} J.~P.,  {Mahadevan} S.,   {HPF Team} 2023, {Evolution of the vertical
  thermal profile of an inner disc in an episodic accretion event},
  \url{http://ppvii.org/chapter/ppvii_abstract/sf06/SF-06-0010.html/}

\bibitem[\protect\citeauthoryear{{Ormel}, {Vazan}  \& {Brouwers}}{{Ormel}
  et~al.}{2021}]{Ormel-21_pebbles_interior}
{Ormel} C.~W.,  {Vazan} A.,   {Brouwers} M.~G.,  2021, \mn@doi [\aap]
  {10.1051/0004-6361/202039706}, \href
  {https://ui.adsabs.harvard.edu/abs/2021A&A...647A.175O} {647, A175}

\bibitem[\protect\citeauthoryear{{Owen} \& {Wu}}{{Owen} \&
  {Wu}}{2016}]{OwenWu-16-boil-off}
{Owen} J.~E.,  {Wu} Y.,  2016, \mn@doi [\apj] {10.3847/0004-637X/817/2/107},
  \href {https://ui.adsabs.harvard.edu/abs/2016ApJ...817..107O} {817, 107}

\bibitem[\protect\citeauthoryear{{Paxton}, {Bildsten}, {Dotter}, {Herwig},
  {Lesaffre}  \& {Timmes}}{{Paxton} et~al.}{2011}]{PaxtonEtal11}
{Paxton} B.,  {Bildsten} L.,  {Dotter} A.,  {Herwig} F.,  {Lesaffre} P.,
  {Timmes} F.,  2011, \mn@doi [\apjs] {10.1088/0067-0049/192/1/3}, \href
  {http://adsabs.harvard.edu/abs/2011ApJS..192....3P} {192, 3}

\bibitem[\protect\citeauthoryear{{Paxton} et~al.,}{{Paxton}
  et~al.}{2013}]{Paxton13-Mesa-2}
{Paxton} B.,  et~al., 2013, \mn@doi [\apjs] {10.1088/0067-0049/208/1/4}, \href
  {https://ui.adsabs.harvard.edu/abs/2013ApJS..208....4P} {208, 4}

\bibitem[\protect\citeauthoryear{{P{\'e}rez} et~al.,}{{P{\'e}rez}
  et~al.}{2020}]{Perez20-FUOri}
{P{\'e}rez} S.,  et~al., 2020, \mn@doi [\apj] {10.3847/1538-4357/ab5c1b}, \href
  {https://ui.adsabs.harvard.edu/abs/2020ApJ...889...59P} {889, 59}

\bibitem[\protect\citeauthoryear{{Powell}, {Irwin}, {Bouvier}  \&
  {Clarke}}{{Powell} et~al.}{2012}]{PowellEtal12}
{Powell} S.~L.,  {Irwin} M.,  {Bouvier} J.,   {Clarke} C.~J.,  2012, \mn@doi
  [\mnras] {10.1111/j.1365-2966.2012.21898.x}, \href
  {http://adsabs.harvard.edu/abs/2012MNRAS.426.3315P} {426, 3315}

\bibitem[\protect\citeauthoryear{{Rodriguez} \& {Hillenbrand}}{{Rodriguez} \&
  {Hillenbrand}}{2022}]{Rodriguez-22-FUORs}
{Rodriguez} A.~C.,  {Hillenbrand} L.~A.,  2022, \mn@doi [\apj]
  {10.3847/1538-4357/ac496b}, \href
  {https://ui.adsabs.harvard.edu/abs/2022ApJ...927..144R} {927, 144}

\bibitem[\protect\citeauthoryear{{Santerne} et~al.,}{{Santerne}
  et~al.}{2016}]{SanterneEtal15}
{Santerne} A.,  et~al., 2016, \mn@doi [\aap] {10.1051/0004-6361/201527329},
  \href {http://adsabs.harvard.edu/abs/2016A%26A...587A..64S} {587, A64}

\bibitem[\protect\citeauthoryear{{Scepi}, {Lesur}, {Dubus}  \& {Flock}}{{Scepi}
  et~al.}{2018}]{Scepi-18-TI-alpha}
{Scepi} N.,  {Lesur} G.,  {Dubus} G.,   {Flock} M.,  2018, \mn@doi [\aap]
  {10.1051/0004-6361/201731900}, \href
  {https://ui.adsabs.harvard.edu/abs/2018A&A...609A..77S} {609, A77}

\bibitem[\protect\citeauthoryear{{Shakura} \& {Sunyaev}}{{Shakura} \&
  {Sunyaev}}{1973}]{Shakura73}
{Shakura} N.~I.,  {Sunyaev} R.~A.,  1973, \aap, \href
  {http://cdsads.u-strasbg.fr/cgi-bin/nph-bib_query?bibcode=1973A%26A....24..337S&db_key=AST}
  {24, 337}

\bibitem[\protect\citeauthoryear{{Siwak} et~al.,}{{Siwak}
  et~al.}{2018}]{Siwak21-FUOri-QPOs}
{Siwak} M.,  et~al., 2018, \mn@doi [\aap] {10.1051/0004-6361/201833401}, \href
  {https://ui.adsabs.harvard.edu/abs/2018A&A...618A..79S} {618, A79}

\bibitem[\protect\citeauthoryear{{Spiegel} \& {Burrows}}{{Spiegel} \&
  {Burrows}}{2012}]{SpiegelBurrows12}
{Spiegel} D.~S.,  {Burrows} A.,  2012, \mn@doi [\apj]
  {10.1088/0004-637X/745/2/174}, \href
  {http://adsabs.harvard.edu/abs/2012ApJ...745..174S} {745, 174}

\bibitem[\protect\citeauthoryear{{Szegedi-Elek} et~al.,}{{Szegedi-Elek}
  et~al.}{2020}]{Szegedi-Elek-20-Gaia-18}
{Szegedi-Elek} E.,  et~al., 2020, \mn@doi [\apj] {10.3847/1538-4357/aba129},
  \href {https://ui.adsabs.harvard.edu/abs/2020ApJ...899..130S} {899, 130}

\bibitem[\protect\citeauthoryear{{Valletta} \& {Helled}}{{Valletta} \&
  {Helled}}{2020}]{VallettaHelled-20-high-Z}
{Valletta} C.,  {Helled} R.,  2020, \mn@doi [\apj] {10.3847/1538-4357/aba904},
  \href {https://ui.adsabs.harvard.edu/abs/2020ApJ...900..133V} {900, 133}

\bibitem[\protect\citeauthoryear{{Vorobyov} \& {Basu}}{{Vorobyov} \&
  {Basu}}{2006}]{VB06}
{Vorobyov} E.~I.,  {Basu} S.,  2006, \mn@doi [\apj] {10.1086/507320}, \href
  {http://adsabs.harvard.edu/abs/2006ApJ...650..956V} {650, 956}

\bibitem[\protect\citeauthoryear{{Vorobyov} \& {Basu}}{{Vorobyov} \&
  {Basu}}{2010}]{VB10}
{Vorobyov} E.~I.,  {Basu} S.,  2010, \mn@doi [\apj]
  {10.1088/0004-637X/719/2/1896}, \href
  {http://adsabs.harvard.edu/abs/2010ApJ...719.1896V} {719, 1896}

\bibitem[\protect\citeauthoryear{{Vorobyov} \& {Basu}}{{Vorobyov} \&
  {Basu}}{2015}]{VB15}
{Vorobyov} E.~I.,  {Basu} S.,  2015, \mn@doi [\apj]
  {10.1088/0004-637X/805/2/115}, \href
  {https://ui.adsabs.harvard.edu/abs/2015ApJ...805..115V} {805, 115}

\bibitem[\protect\citeauthoryear{{Vorobyov} \& {Elbakyan}}{{Vorobyov} \&
  {Elbakyan}}{2019}]{Vorobyov-Elbakyan-19}
{Vorobyov} E.~I.,  {Elbakyan} V.~G.,  2019, arXiv e-prints, \href
  {https://ui.adsabs.harvard.edu/abs/2019arXiv190810589V} {p. arXiv:1908.10589}

\bibitem[\protect\citeauthoryear{{Vorobyov}, {Elbakyan}, {Liu}  \&
  {Takami}}{{Vorobyov} et~al.}{2021}]{Vorobyov-21-flyby-FUORs}
{Vorobyov} E.~I.,  {Elbakyan} V.~G.,  {Liu} H.~B.,   {Takami} M.,  2021,
  \mn@doi [\aap] {10.1051/0004-6361/202039391}, \href
  {https://ui.adsabs.harvard.edu/abs/2021A&A...647A..44V} {647, A44}

\bibitem[\protect\citeauthoryear{{Wuchterl}, {Guillot}  \&
  {Lissauer}}{{Wuchterl} et~al.}{2000}]{WuchterlEtal00}
{Wuchterl} G.,  {Guillot} T.,   {Lissauer} J.~J.,  2000, Protostars and Planets
  IV, \href {http://ukads.nottingham.ac.uk/abs/2000prpl.conf.1081W} {pp
  1081--+}

\bibitem[\protect\citeauthoryear{{Zakri} et~al.,}{{Zakri}
  et~al.}{2022}]{Zakri-22-FUOR-Class-0}
{Zakri} W.,  et~al., 2022, \mn@doi [\apjl] {10.3847/2041-8213/ac46ae}, \href
  {https://ui.adsabs.harvard.edu/abs/2022ApJ...924L..23Z} {924, L23}

\bibitem[\protect\citeauthoryear{{Zhu}, {Hartmann}, {Calvet}, {Hernandez},
  {Muzerolle}  \& {Tannirkulam}}{{Zhu} et~al.}{2007}]{ZhuEtal07}
{Zhu} Z.,  {Hartmann} L.,  {Calvet} N.,  {Hernandez} J.,  {Muzerolle} J.,
  {Tannirkulam} A.-K.,  2007, \mn@doi [\apj] {10.1086/521345}, \href
  {http://adsabs.harvard.edu/abs/2007ApJ...669..483Z} {669, 483}

\bibitem[\protect\citeauthoryear{{Zhu}, {Hartmann}, {Calvet}, {Hernandez},
  {Tannirkulam}  \& {D'Alessio}}{{Zhu} et~al.}{2008}]{Zhu08-FUOri}
{Zhu} Z.,  {Hartmann} L.,  {Calvet} N.,  {Hernandez} J.,  {Tannirkulam} A.-K.,
   {D'Alessio} P.,  2008, \mn@doi [\apj] {10.1086/590241}, \href
  {https://ui.adsabs.harvard.edu/abs/2008ApJ...684.1281Z} {684, 1281}

\bibitem[\protect\citeauthoryear{{Zhu}, {Hartmann}  \& {Gammie}}{{Zhu}
  et~al.}{2009a}]{ZhuEtal09}
{Zhu} Z.,  {Hartmann} L.,   {Gammie} C.,  2009a, \mn@doi [\apj]
  {10.1088/0004-637X/694/2/1045}, \href
  {http://adsabs.harvard.edu/abs/2009ApJ...694.1045Z} {694, 1045}

\bibitem[\protect\citeauthoryear{{Zhu}, {Espaillat}, {Hinkle}, {Hernandez},
  {Hartmann}  \& {Calvet}}{{Zhu} et~al.}{2009b}]{Zhu09-FUOri-obs}
{Zhu} Z.,  {Espaillat} C.,  {Hinkle} K.,  {Hernandez} J.,  {Hartmann} L.,
  {Calvet} N.,  2009b, \mn@doi [\apjl] {10.1088/0004-637X/694/1/L64}, \href
  {https://ui.adsabs.harvard.edu/abs/2009ApJ...694L..64Z} {694, L64}

\bibitem[\protect\citeauthoryear{{Zhu}, {Hartmann}, {Gammie}  \&
  {McKinney}}{{Zhu} et~al.}{2009c}]{Zhu09-FUOR-2D-MRI-sims}
{Zhu} Z.,  {Hartmann} L.,  {Gammie} C.,   {McKinney} J.~C.,  2009c, \mn@doi
  [\apj] {10.1088/0004-637X/701/1/620}, \href
  {https://ui.adsabs.harvard.edu/abs/2009ApJ...701..620Z} {701, 620}

\bibitem[\protect\citeauthoryear{{Zhu}, {Hartmann}  \& {Gammie}}{{Zhu}
  et~al.}{2010}]{Zhu10-DZ-MRI-1D}
{Zhu} Z.,  {Hartmann} L.,   {Gammie} C.,  2010, \mn@doi [\apj]
  {10.1088/0004-637X/713/2/1143}, \href
  {https://ui.adsabs.harvard.edu/abs/2010ApJ...713.1143Z} {713, 1143}

\makeatother
\end{thebibliography}

\bsp	
\label{lastpage}
\end{document}